\RequirePackage{fix-cm}
\documentclass[twocolumn]{svjour3}          
\smartqed  
\usepackage{graphicx}
\usepackage{latexsym,amscd,amssymb,amsopn,amsfonts,amsmath}

\usepackage{color}
\usepackage{cite}
\usepackage{bm}
\usepackage{amsfonts}
\DeclareMathOperator{\sgn}{sgn}

\begin{document}

\title{Aspects of topological superconductivity in 2D systems: noncollinear magnetism, skyrmions, and higher-order topology
\thanks{The study was funded by the Russian Foundation for Basic Research (Project No. 19-02-00348), Government of Krasnoyarsk Territory, Krasnoyarsk Regional Fund of Science (Grants Nos. 19-42-240011, 20-42-243001). A.O.Z. and M.S.S. are grateful to the support of the Foundation for the Advancement of Theoretical Physics and Mathematics ``BASIS''.}
}

\titlerunning{Aspects of topological superconductivity in 2D systems}        

\author{A.O. Zlotnikov         \and
        M.S. Shustin \and
        A.D. Fedoseev 
}


\institute{A.O. Zlotnikov \at
              Kirensky Institute of Physics, Federal Research Center KSC SB RAS, 660036 Krasnoyarsk, Russia\\
              Tel.: +7-983-159-90-44\\
              Fax: +7-391-243-89-23\\
              \email{zlotn@iph.krasn.ru}           
           \and
           M.S. Shustin \at
              Kirensky Institute of Physics, Federal Research Center KSC SB RAS, 660036 Krasnoyarsk, Russia
           \and
           A.D. Fedoseev \at
              Kirensky Institute of Physics, Federal Research Center KSC SB RAS, 660036 Krasnoyarsk, Russia
}

\date{Received: date / Accepted: date}

\maketitle

\begin{abstract}

The review is aimed at highlighting the aspects of topological superconductivity in the absence of spin-orbit interaction in two-dimensional systems with long-range non-collinear spin ordering or magnetic skyr\-mi\-ons. Another purpose is to give a brief introduction to the new concept of topological superconductivity, i.e. higher-order topology in two-dimensional systems including spin-orbit coupled structures. The formation of Majorana modes due to magnetic textures is discussed. The role of effective triplet pairings and odd fermion parity of the ground state wave function in different systems is emphasized. We describe the peculiarities of the magnetic skyrmions, leading to the formation of the Majorana modes and defects on which the modes are localized. The problem of braiding in the two-dimensional systems, especially in higher-order topological superconductors, is considered.
\keywords{Topological superconductivity \and Majorana fermions \and Noncollinear magnetism \and Magnetic skyrmion \and Higher-order topology}
\PACS{74.70.-b \and 03.65.Vf \and 12.39.Dc \and 75.50.-y}
\end{abstract}

\section{Introduction}
\label{intro}

Until recently, the actively studied quantum phases of condensed matter were homeomorphic to the vacuum state: a continuous change in the parameters of the system could reduce the latter into a classical vacuum state, corresponding to a set of non-interacting and non-entangled atoms. However, it turned out that there are quantum phases that cannot be adiabatically deformed to a vacuum state without violating the certain continuity conditions (overcoming a high energy barrier, closing a gap in the excitation spectrum, etc.).  The discovery of such quantum states raised the question of the classification of continuous mappings between them.  Since the mathematical side of such a problem belongs to the field of topology, the described phases were called topological phases. So, two quantum states belonging to different topological phases (or, mathematically, homotopy classes) of the same system cannot be continuously deformed into each other. In most practically encountered topological systems, homotopy classes form Abelian groups isomorphic to either the group $\mathbb{Z}$ of integers or to the groups $\mathbb{Z}_n$ of integers modulo $ n $. In this case, the correspondence of a certain homotopy class to an integer can be obtained by calculating the topological invariant. The calculation of the latter depends on the details of the system: its symmetry, dimension, absence or presence of interactions, etc. Condensed matter systems hosting different topological phases for which the continuity conditions imply the preservation of the excitation gap are called topological insulators or superconductors. In this case, the vacuum state is usually belonged to a topologically trivial class, usually corresponding to the zero value of the topological invariant. The other values of topological invariant are called non-trivial.

The new classes of insulators and superconductors were suggested over 20 years ago. The first topological insulator is the quantum spin Hall insulator \cite{kane-05}, while the initial topological superconductors (TSCs) are chiral superconductors (d-wave \cite{volovik-97} or p-wave \cite{read-00}). It is essential that topological insulators and superconductors have the gapped excitation spectrum in the homogeneous case and with the periodic boundary conditions (such spectrum is called the bulk spectrum). At the same time the topological invariant calculated for the homogeneous case can have nontrivial values. According to the bulk-boundary correspondence for topological phases, inhomogeneous objects such as edges, Abrikosov vortex, domain walls, etc., lead not only to the quantitative changes of the spectral features, but also to the qualitative new solutions as an appearance of subgap states. Moreover, it is believed that such excitations are topologically protected that means that they are stable unless the bulk spectrum is gapped or the existing symmetry is preserved. When the bulk spectrum becomes gapless the topological phase transition occurs with a change of the topological invariant. It can be transition to the trivial phase where the subgap states are absent or they are not topologically protected. On the other hand,
the topological transition to the phase with other topological invariant can exist.

It is known from the quantum Hall effect theory that the quantization of the Hall conductivity is determined by an integer number which is expressed in the term of the first Chern number \cite{kohmoto-85}. In the simple cases of 2D chiral p-wave and d-wave superconductors, the Chern number has a sense of the winding number of the pseudospin vector $\left(\Re(\Delta_k), -\Im(\Delta_k), \xi_k \right)/\varepsilon_k$ and means the number of times this vector winds upon running over the Brillouin zone \cite{read-00}. Such pseudospin vector was introduced by Anderson in the mean-field description of the superconducting state \cite{anderson-58}.

It should be noted that not all TSCs support so-called Majorana bound states. For example, topologically protected non-Majorana edge sta\-tes appear in spin-singlet chiral d-wave superconductors. Nevertheless, the main interest in topological superconductivity (TSCty) is based on Majorana modes (MMs).

MMs as topological excitations in low-dimensional TSCs were predicted about 20 years ago for two-di\-men\-sio\-nal (2D) \cite{read-00, ivanov-01} and one-di\-men\-sio\-nal (1D) \cite{kitaev-01} non-interacting electronic systems with nontrivial superconducting pairing. The key ingredients of the proposed models are the spin-triplet chiral $p_x+ip_y$ pairing in the 2D case \cite{read-00} or spinless pairing for the model of the Kitaev chain \cite{kitaev-01} which cause MMs formation. In general, all proposals of TSCs with MMs assume the existence of the effective triplet superconductivity.

MMs are characterized by zero excitation energy and spatial nonlocality, i.e. their wave functions should not overlap, and they are always created in pairs. To describe MMs in the case of a doubly degenerate ground state two Majorana operators $b^{\prime}$ and $b^{\prime \prime}$ (following Kitaev \cite{kitaev-01}) can be constructed as two linearly independent superpositions of Bogoliubov creation and annihilation quasiparticle operators corresponding to an excitation with zero energy. The operators $b^{\prime}$ and $b^{\prime \prime}$ are self-conjugate $(b^{\prime (\prime \prime)})^{\dag} = b^{\prime (\prime \prime)}$ and satisfy the commutation relations $\{b^{\prime (\prime \prime)}, b^{\prime (\prime \prime)}\} = 2$, $\{b^{\prime}, b^{\prime \prime}\} = 0$. Bearing it in mind, MMs in condensed matter systems are also called as Majorana fermions, Majorana excitations, or Majorana quasiparticles. Nevertheless, it should be noted that such fermions are significantly distinct from Bogoliubov quasiparticles in solid state physics and Majorana fermions in particle physics. The existence of two MMs with non-overlapping wave functions is often interpreted as the realization of one Majorana bound state (MBS).

In quasi-1D quantum wires, MMs can be localized at the edges or inhomogeneities \cite{kaladzhyan-19}, while in 2D systems they can occur on superconducting \cite{read-00, bjornson-13, akzyanov-16, akzyanov-17} or magnetic vortices \cite{rex-19} and corners \cite{wang-18b} (see, also references below). In this case, the nonlocal structure of Majorana quasiparticles is the reason for their stability to local external perturbations. Therefore, this feature is often referred to as the topological stability of MBS. Moreover, Majorana excitations are examples of anions, quasi-particles with non-Bose or non-Fermi statistics. Therefore, they are proposed for the realization of topologically protected quantum computations through the exchange of the positions of two Majorana excitations in the system which is also called braiding procedure \cite{nayak-08}.  These features give rise to a considerable interest in solid-state systems with MMs as promising materials for quantum computing devices.

At present, TSCty has become an extensive and intensively developed area of condensed matter physics \cite{alicea-12, beenakker-13, elliott-15, sato-17, usp-21, samokhvalov-21}. For example, superconducting pairing of electrons with the same spin projections (triplet pairing) was found to play an important role in the realization of TSCty with MMs. Since there are quite a few candidates for triplet superconductors (Sr$_2$RuO$_4$, UGe$_2$, UCoGe, URhGe \cite{mackenzie-03, das_sarma-06, sau-12, mineev-17, pustogow-19, suzuki-20}), it turns out to be important to search for the conditions for MMs in other compounds and structures. To date, the most studied TSCs are:
\begin{enumerate}
    \item hybrid structures of a conventional superconductor and a semiconductor with the strong spin-orbit interaction \cite{sau-10, lutchyn-10, oreg-10, deng-12};

    \item systems with coexisting spin-singlet superconductivity and noncollinear magnetic ordering: either materials with the homogeneous coexistence phase \cite{martin-12, lu-13}, or heterostructures of superconducting and magnetic layers \cite{choy-11, chen-15};

    \item superconductor / chiral magnet hybrids with magnetic skyrmion hybrid structures. In this case, one of MMs is localized around the magnetic skyrmion (MS) \cite{yang-16, rex-19, mascot-21, mohanta-21}. Actually, the developed technologies for controllable movement of MSs essentially allow braiding of Majorana modes and creating stable qubits by using MSs;

    \item higher-order topological superconductors (HOTSCs) \cite{Zhang-20-prr,Zhang-20-prb,Wu-20,Zhang-19,Zhu-18}. In such systems, bulk and ordinary edge state spectra are gapped, but there are lattice-free edge states of higher order. These states including MMs are localized at the corners of 2D systems or at the hinges, vertices, and other topological defects of 3D systems. It should be noted that the existence of zero modes on the line on the surface was earlier shown in the topological superfluid $^{3}$He-B in magnetic field \cite{volovik-10};

    \item heterostructures of topological insulator / superconductor and doped topological insulators in which superconductivity occurs, such as Cu$_x$Bi$_2$Se$_3$ \cite{fu-08, hor-10, hirsch-15}. This group may also include iron-based superconductors such as FeTe$_x$Se$_{1-x}$ \cite{zhang-18} in which topological surface states are formed in the normal non-superconducting phase.
\end{enumerate}
All the mentioned systems contain the necessary mechanism which transforms conventional spin-singlet superconductivity to TSCty. It is known that spin degrees of freedom must be mixed in a spin-singlet superconductor to support the topological order.

Previously, it was suggested that the spin-orbit interaction in a uniform magnetic field should fulfil the role of this mechanism and induce MMs.
The main experimental progress has also been achieved for quantum wires with spin-orbit coupling (the first group in the list above). In particular, the transport properties of InAs or InSb semiconductor nanowires contacting with a superconducting electrode (NbTiN) either partially or fully covered with a thin Al or Nb layer were studied \cite{mourik-12, deng-12, deng-16, nichele-17, vaitiekenas-20, shen-21}. A stable zero-bias peak (ZBP) in conductance was observed. The appearance of this peak was associated with the implementation of a topologically nontrivial phase and MMs, although experimental and theoretical discussions on this issue continue (see, for example, Refs.\cite{moore-18, reeg-18, zhang-21}). In addition, hybrid structures Al-EuS-InAs with a thin ferromagnetic EuS layer were synthesized and experimentally studied \cite{vaitiekenas-21}. The proximity-induced ferromagnetic correlations due to the EuS layer made it possible to observe a quantized peak of differential conductance in the absence of external magnetic fields. The qualitatively same systems are heterostructures with a spin-orbit coupled superconductor (such as Pb) and a ferromagnetic layer or nanoisland \cite{li-16}. Scanning tunneling spectroscopy experiments were carried out for such materials and features of MMs were found \cite{menard-19}.

Note that the studies of superconducting nanowires revealed a number of problems inherent of 1D TSCs: the need to create complex hybrid structures for experimental research \cite{lutchyn-18}; few suitable candidate materials for TSC; the need to create $T-$, $X-$ and $Y-$junctions for MM braiding \cite{Alicea-11}.

Most of the detailed reviews concerning the topological superconductivity and MMs \cite{alicea-12, beenakker-13, elliott-15, sato-17, usp-21, samokhvalov-21} are devoted to topological systems with spin-orbit coupling (SOC). In this review, we will discuss the aspects of 2D TSCty where the topological order appears when SOC can be neglected (namely, points 2 and 3 from the list above) and 2D HOTSCs for which there is another component in addition to the spin-orbital one for inducing strongly localized MMs in the 2D case.

While the main properties of MMs were well described in 1D systems, there are some features which exist only in 2D systems. Firstly, the conventional 2D TSC has a gapless edge excitation spectrum in the open cylinder (or strip) geometry (open boundaries in one direction and periodic boundary conditions in the other) instead of several zero energy solutions in the 1D case. MMs in the 2D case are still characterized by zero energy and spatial nonlocality, with these MMs being localized at the opposite edges, but propagating along these edges. Secondly, in the 2D case, MMs can be localized on the defects and their position can be controlled. The Abrikosov vortex created under a magnetic field or magnetic inhomogeneities including magnetic skyrmion can play the role of such a defect. This allows one to manipulate the position of MMs and consequently, can provide braiding. Thirdly, the higher system dimension opens the possibility for a new class of topological material: HOTSCs, which are obviously impossible in the 1D case. While HOTSCs provide point-localized MMs in the 2D environment, they are also attractive systems to provide braiding. The review of the most promising features for creating MMs in 2D is the main objective of the present paper.

The review consists of three independent parts. Section \ref{sec:2} is devoted to the TSCty induced by long-range noncollinear spin ordering in the absence of the pronounced spin-orbit interaction and uniform magnetic field. The general idea and its implementation in the 2D case are discussed. Section \ref{sec:3} deals with the concept of magnetic skyrmions in TSC and their advantages for the realization of MMs. The HOTSC concept, list of the most promising models, examples of braiding processes on HOTSC are presented in Section \ref{sec:4}.

\section{Topological superconductivity due to noncollinear spin ordering}
\label{sec:2}

\subsection{1D case}\label{sec:2.1}

As soon as the first theoretical proposals revealed the role of the spin-orbit interaction in TSCs, it was shown \cite{braunecker-10} that the 1D Hamiltonian with the Rashba spin-orbit interaction in the presence of the magnetic field perpendicular to the spin-orbit field vector is connected by the unitary transformation with the Hamiltonian having the noncollinear magnetic field
\begin{equation}
{\bf h}_f = h \left( \cos(Q R_f), -\sin(Q R_f), 0 \right).
\end{equation}
Here, $h$ is the Zeeman energy in the initial Hamiltonian and the wave-number $Q$ determining the period of a noncollinear structure is associated with the Rashba spin-orbit coupling parameter $\alpha$ by the formula
\begin{equation}
\label{Qa}
Qa = 2 \arccos\left( \left( \sqrt{ 1 + \alpha^2/t^2 } \right)^{-1} \right),
\end{equation}
where $a$ is the chain parameter and $t$ is the hopping amplitude of electrons between the nearest sites of the chain. A more detailed description of the correspondence between the above mentioned systems can be found in \cite{usp-21}.

In general, noncollinear magnetic structures include a spin helix, spin spiral, cycloidal texture, etc., which will be considered later. We will assume only magnetic orderings which are commensurate with the lattice period.

From (\ref{Qa}) it is seen that the ferromagnetic ordering with $Q=0$ corresponds to the trivial case with $\alpha = 0$, while the antiferromagnetic order with $Q=\pi/a$ cannot be realized due to the spin-orbit interaction since it corresponds to $\alpha/t \to \infty$. Indeed,  Majorana fermions were proposed to exist in an antiferromagnetic chain only in the presence of the weak Zeeman field and a supercurrent induced in a superconducting substrate \cite{heimes-14}. For other values of the amplitude $\alpha$ different noncollinear magnetic textures appear. For example, the 120-degree (120$^{\circ}$) magnetic structure $Q=2\pi/3a$ is formed for $\alpha = \sqrt{3}t$.

\begin{figure}
\begin{center}
  \includegraphics{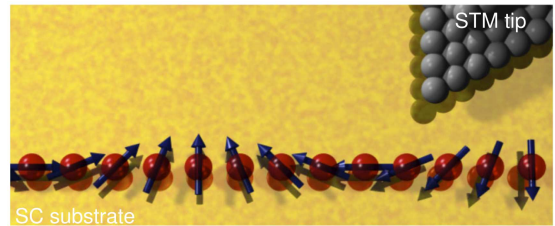}
\caption{Chain of magnetic atoms on a superconducting
substrate  \cite{vazifeh-13}.}
\label{fig_vazifeh_13}
\end{center}
\end{figure}

\begin{figure}
\begin{center}
  \includegraphics{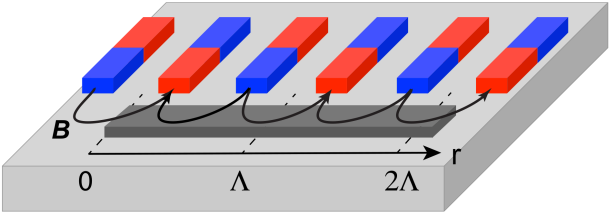}
\caption{The nanowire (dark grey stripe) is in the proximity of the superconductor
(grey rectangle) in the vicinity of the nanomagnets (blue-red rectangles) \cite{kornich-20}.}
\label{fig_kornich_20}
\end{center}
\end{figure}

Using the generic idea in  \cite{braunecker-10} different candidates for TSCs are suggested, such as
\begin{itemize}
  \item a chain of magnetic nanoparticles or adatoms with an arbitrary direction of magnetic moments which is deposited on a superconducting substrate \cite{choy-11, vazifeh-13, nadj-perge-13, pientka-13, schecter-16, christensen-16} (see fig. {\ref{fig_vazifeh_13}}). It is believed that the spiral magnetic ordering can be caused by the substrate;
  \item a superconducting wire \cite{kjaergaard-12, egger-12, klinovaja-12, kornich-20} in a helical magnetic field created by submicron magnets periodically located in the vicinity of the wire (see fig. {\ref{fig_kornich_20}});
  \item different wires with the proximity-induced superconductivity in which the noncollinear magnetic ordering is caused by internal interactions, e. g. the RKKY interaction of localized electrons or nuclear spins induced by conduction electrons of the wire \cite{braunecker-13, klinovaja-13, hsu-15}.
\end{itemize}

It should be noted that the chain of magnetic atoms in such structures can be dense \cite{nadj-perge-13} when the orbitals of different atoms are strongly overlapped or it can be dilute forming subgap Shiba bands due to the hybridization of the Shiba bound states of different magnetic impurities \cite{pientka-13}. Both systems support the formation of Majorana modes (MMs). In the case of the spin Shiba chain, the hopping and pairing terms have the $1/r$ power law decay at the distance $r\ll\xi$, where $\xi$ is the coherence length, and complex phase factors of hopping terms. Such features lead to the power-law dependence with logarithmic corrections of the spatial decay of two MMs on the chain length instead of the exponential ones in the case of the dense wire.

These suggestions motivate tunneling spectroscopy experiments on Fe chains on bulk superconducting Pb \cite{nadj-perge-14, ruby-15, feldman-17}. In these experiments the zero bias peak (ZBP) in the conductance is demonstrated when the microscope tip is located near the edges of the chain which is attributed to MBS. Nevertheless, ZBP is not quantized to the predicted value $2e^2/h$ for MMs. Moreover, it is supposed that ordering in the Fe chains is ferromagnetic and ZBP could be caused by the spin-orbit interaction in Pb \cite{nadj-perge-14, peng-15}. This conclusion is consistent with the results in Ref. \cite{kim-14} where it is shown that the 1D spin helix state becomes unstable in the presence of disorder. Moreover, to stabilize the helix structure a sufficiently strong Rashba SOC can be required.

It is shown that the noncollinear 120$^{\circ}$ spin ordering in the Fe chain can be realized when the chain is deposited on Ir(001) \cite{menzel-12} or Re(0001) \cite{kim-18} surfaces. It is supposed that the Dzyaloshinskii-Moriya interaction plays an essential role in such structures. For the last structure, Re becomes superconducting at low temperature. Therefore, experiments on scanning tunneling microscopy were carried out to observe MBS \cite{kim-18}. ZBP was not observed and only the local maximum of the conductance at the zero bias is shown. This result cannot be considered as clear evidence for the MBS formation. Nevertheless, these structures are perspective for different Majorana devices.

Experimentally, the helical magnetic field is obtained for a single-walled carbon nanotube with superconducting electrodes coupled to a magnetically textured gate \cite{desjardins-19} and a hybrid semiconductor-superconductor nano\-wire on the top of a magnetic film in the stripe phase \cite{mohanta-19}. From the observed oscillations of conductance under the magnetic field the presence of the synthetic spin-orbit interaction is concluded due to the inhomogeneous magnetic field \cite{desjardins-19}. A tiny ZBP is also observed.

It should be noted that the Majorana fermions are realized not only on spin chains but also on a ladder of two or more coupled chains \cite{poyhonen-14, xiao-15}. An interesting result is that the topologically trivial chains coupled by electron hopping may form a topologically nontrivial ladder supporting the Majorana end states.

In a certain sense, the formation of the noncollinear spin ordering in the 2D and quasi-2D structures could be easier achieved than in the 1D case. Therefore, in the next section we will discuss the problem of the MBS appearance in the 2D structures with the coexisting superconductivity and noncollinear spin ordering.

\subsection{2D case} \label{sec:2.2}
We will start with the general view of the Hamiltonian describing in the mean-field approximation the coexistence of superconductivity and noncollinear, but coplanar spin ordering in different 2D structures and materials:
\begin{eqnarray}
\label{Ham}
H & = & -\mu \sum_{f \sigma} c_{f \sigma}^{\dag}c_{f \sigma} + \sum_{fm \sigma} t_{fm} c_{f \sigma}^{\dag}c_{m \sigma} +
\nonumber \\
& + & \sum_{f} \left( h_f ({\bf Q}) e^{i{\bf QR}_f} c_{f \uparrow}^{\dag}c_{f \downarrow} + H.c.  \right) +
\nonumber \\
& + & \sum_{fm} \left( \Delta_{fm}  c^{\dag}_{f \uparrow}c^{\dag}_{m \downarrow} + H.c.  \right),
\end{eqnarray}
where $\mu$ is the chemical potential, $t_{fm}$ is the hopping parameter, $h_f({\bf Q})$ is the exchange field, depending on the spin structure vector ${\bf Q}$, such as the vector spin operator defined as
\begin{equation}
\label{Sf}
\left\langle {\bf S}_f \right\rangle = M_f \left( \cos({\bf QR}_f), -\sin({\bf QR}_f), 0 \right).
\end{equation}
To be specific, we consider the vectors ${\bf Q}$ of the noncollinear spin structures with the components  $Q_i \in (0, \pi)$ in a chosen basis. The indices $f$ and $m$ denote the radius-vectors ${\bf R}_f$ and ${\bf R}_m$ of the lattice sites. The parameter $\Delta_{fm}$ defines the amplitude of superconducting pairings between the fermions on the same site ($\Delta_{fm} = \Delta \delta_{fm}$, where $\delta_{fm}$ is the Kronecker symbol) or different ($f \ne m$) lattice sites. The exchange field is described by the expression
\begin{equation}
\label{h_Thor}
h_f\left(\bf{Q}\right) = \frac{1}{2} \sum_{m}J_{fm} M_m \exp(-i {\bf Q}({\bf R}_f-{\bf R}_m)), \nonumber
\end{equation}
where $J_{fm}$ is the parameter of the exchange interaction between the electrons. In the mean-field approximation the parameter $J_{fm}$ can correspond to the \emph{s-d} exchange interaction between the itinerant electrons and localized spins or the exchange interaction between the same itinerant electrons. It can be the local on-site parameter ($J_{fm} = J\delta_{fm}$) or it depends on the distance between the lattice sites.

There are different mechanisms causing noncollinear spin ordering. Among them are geometric frustrations, for example, as can be observed in a triangular lattice; competition of the exchange bonds of a spin with the neighboring spins from different coordination spheres (as an example \cite{mikheenkov-18}); influence of superconductivity \cite{devizorova-19}, etc. At the mean-field level we will not discuss the specific mechanism.

For simplicity, we limit the consideration to the case when the spin ordering is realized in a plane of the spin space in the absence of an external magnetic field. Nevertheless, a more complex case in the problem of topological superconductivity is often considered \cite{choy-11, nakosai-13, egger-12} with the average spin operator
\begin{equation}
\label{GSf}
\left\langle {\bf S}_f \right\rangle = M_f \left( \sin(\phi_f)\cos(\theta_f), \sin(\phi_f)\sin(\theta_f), \cos(\phi_f) \right).
\end{equation}

It should be noted that it is convenient to use the definition (\ref{Sf}) for the description of the noncollinear ordering instead of using many-sublattice representation. Actually, in the momentum representation the space of the electron states is limited by the subspace of the states $(k, \sigma = \uparrow)$ and $(k-Q, \sigma = \downarrow)$ invariant under the action of Hamiltonian for all $k$ \cite{igoshev-10}. Therefore, the calculations for the whole first Brillouin zone (BZ) in the nonmagnetic case must be carried out with the periodic boundary conditions \cite{chubukov-94, igoshev-10}. For a more complex case (\ref{GSf}) the magnetic BZ must be considered. The topological classification taking into account magnetic groups was made in \cite{steffensen-20}.

\begin{figure}
\begin{center}
\includegraphics[width=0.4\textwidth]{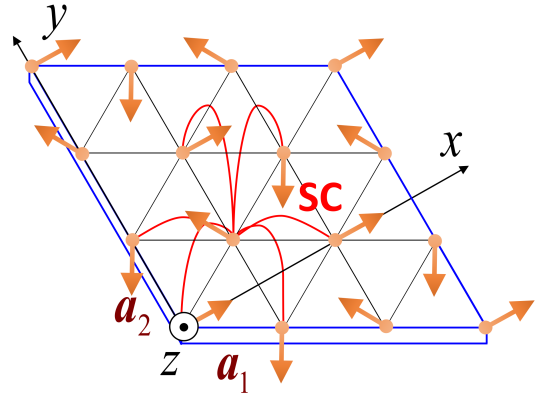}
\caption{The sketch of a superconductor (SC) with the triangular lattice and 120$^{\circ}$ spin ordering in the chosen coordinate frame $xyz$ coinciding with the one in the spin space (see (\ref{Sf})). The arrows denote the spin moments lying in the $xy$ plane of the electrons on the lattice sites. The red lines denote hoppings and pairings between the electrons on the nearest sites which exist at any site. ${\bf a}_1$, ${\bf a}_2$ are the primitive vectors of the triangular lattice.}
\label{fig_sc_120}
\end{center}
\end{figure}

In general, mean-field models similar to (\ref{Ham}) have been widely used to describe the TSCty and MMs in the 2D systems with noncollinear magnetism. Different suggestions can be divided into two main classes. The first class includes magnetic superconductors in which both superconducting and magnetic orderings are caused by intrinsic interactions, for example
\begin{itemize}
  \item helical magnetic s-wave superconductors with a squa\-re lattice \cite{martin-12}. It is believed that this is the case of ternary rare-earth borides or chalcogenides HoMo$_6$S$_8$ and ErRh$_4$B$_4$ \cite{buzdin-84}. These materials have different electron subsystems, localized and itinerant, which are independently responsible for the noncollinear spin order and Cooper instability, respectively. The coupling between different electrons is usually described by the local \emph{s-d} exchange interaction;
  \item triangular lattice superconductors with the chiral d-wave symmetry of the superconducting order parameter and stripe \cite{lu-13} or 120$^{\circ}$ \cite{VVV-ZAO-FAD-ShMS-17, VVV-ZAO-ShMS-18} magnetic ordering (see Fig. \ref{fig_sc_120}). It is supposed that the same electrons can be responsible for the coexistence state in this case. The chiral symmetry of superconductivity is supported by the symmetry of the triangular lattice \cite{zhou-08};
  \item iron-based superconductors with the coexistence of the multi-band spin-singlet superconductivity and different magnetic textures (such as helix, spin whirl, skyrmion in an external magnetic field) \cite{steffensen-20}. In Ref. \cite{steffensen-20} two-band models for iron-based superconductors are studied. Recently, the coexistence phase of superconductivity with the \emph{s}$_{\pm}$ pairing symmetry and helical magnetic ordering along the $c$ axis was found in EuRbFe$_4$As$_4$ \cite{iida-19}. Nevertheless, the interplay between superconductivity and magnetism in this compound is still under debate \cite{kim-21, collomb-21}.
\end{itemize}

The second class consists of hybrid structures with superconducting and magnetic layers. In this case the Hamiltonian (\ref{Ham}) describes the interface of the superconductor and chiral magnet where superconductivity or magnetic ordering are induced due to the proximity effect. Among such heterostructures are

\begin{itemize}
  \item 2D superconducting systems in the helical magnetic field \cite{klinovaja-13a, sedlmayr-15};
  \item magnet-superconductor hybrid systems \cite{chen-15} including the case of a triple-$Q$ magnetic structure \cite{bedow-20} and helical or cycloidal textures \cite{rex-20} (120$^{\circ}$ magnetic ordering from Fig. \ref{fig_sc_120} can also be included along with the proximity induced s-wave superconductivity). A nanoscale Fe monolayer on the oxygen-re\-con\-struc\-ted
  surface of the s-wave superconductor Re(0001) (further, referred to as the Fe/Re structure) was experimentally grown \cite{palacio-morales-16, palacio-morales-19}. It was shown by spin-polarized scanning tunneling micro\-scopy that the 120$^{\circ}$ in-plane spin ordering can be realized in the Fe/Re structure \cite{palacio-morales-16}. Some features of ZBP on the edges of the Fe layer was obtained in \cite{palacio-morales-19}. The triple-$Q$ magnetic structure was experimentally observed in the Mn/Re hybrid \cite{spethmann-20};
  \item a two-dimensional electron gas
formed in a semiconductor quantum well grown on the surface of an s-wave superconductor with a nearby
array of magnetic tunnel junctions \cite{fatin-16, zhou-19}.
\end{itemize}

All the mentioned systems support the Majorana end states in a certain range of model parameters. It should be noted that for some magnetic textures the presence of initial spin-orbit coupling is necessary for the formation of the topologically nontrivial phase. Nevertheless, the magnetic ordering can lead to nontrivial results. For example, the Majorana bound states in such systems can exist not only on the edges but also on disclination defects and domain walls in the magnetic texture \cite{poyhonen-14, rex-20}.

The background reasons of the formation of TSCty in superconductors with helical magnetic ordering can be considered at the qualitative level. Firstly, there is a certain analogy between the noncollinear spin ordering without an external field and SOC in the presence of the magnetic field. Secondly, the features of spin band filling may cause the formation of effective triplet superconductivity due to the noncollinear spin ordering.

To illustrate, use is often made of the unitary transformation to the rotation coordinate frame, by which $\left\langle {\bf S}_f \right\rangle$ becomes aligned with the new $z$ axis at each site \cite{choy-11, egger-12, chen-15}:
\begin{eqnarray}
\label{U}
H \to \tilde{H} & = & U H U^{\dag}, \\ \nonumber U & = & \prod_f
\left[ \exp\left(-i \sgn (h) \frac{\pi}{2}S_f^y\right) \exp\left(-i {\bf Q R}_f
S_f^z \right) \right].
\end{eqnarray}
Fulfilling the periodic boundary conditions and assuming that the exchange field parameter $h_f$ is site-inde\-pen\-dent, the transformed Hamiltonian in the momentum space has the form
\begin{eqnarray}
\label{Ham_K}
&&\tilde{H} = \sum_{k \sigma} \left( \xi_k^{+} - \eta_{\sigma}\left|h\right| \right) c_{k \sigma}^{\dag}c_{k \sigma} -
\sgn (h) \sum_{k \sigma} t_k^- c_{k \sigma}^{\dag}c_{k \bar{\sigma}}
\nonumber \\
&+& \sum_{k} \left[ \Delta_{k}^+  c^{\dag}_{k \uparrow}c^{\dag}_{-k \downarrow} +  \frac{\sgn(h) \Delta_{k}^-}{2} \left( c^{\dag}_{k \uparrow}c^{\dag}_{-k \uparrow} - c^{\dag}_{k \downarrow}c^{\dag}_{-k \downarrow}  \right) + \right.
\nonumber \\
&+& \left. H.c.  \right].
\end{eqnarray}
Here, the function $\eta_{\sigma}$ is +1 for $\sigma = \uparrow$ and -1, otherwise for $\sigma = \downarrow$; $\sgn (h)$ is the signum function $\sgn (h) = 1$ for $h>0$, and  $\sgn (h) = -1$ for $h<0$; $\bar{\sigma}$ denotes the opposite direction of the spin moment $\sigma$. The following notations are introduced:
\begin{eqnarray}
\xi_k^{+}({\bf Q}) & = & t_k^{+}({\bf Q}) - \mu,
\\
t_k^{\pm}({\bf Q}) & = & \frac{1}{2}\left( t_{k-Q/2} \pm t_{k+Q/2} \right),
\\
\Delta_k^{\pm}({\bf Q}) & = & \frac{1}{2}\left( \Delta_{k-Q/2} \pm \Delta_{k+Q/2} \right).
\end{eqnarray}
It is seen that after the transformation the effective triplet pairings are induced in the Hamiltonian for non\-ze\-ro ${\bf Q}$, as it was previously predicted for the antiferromagnetic case \cite{psaltakis-83, kyung-00}. The amplitude of the triplet pairings is described by the odd function $\Delta_k^{-}$ of the quasi-momentum. It should be noted that the effective triplet pairings appear only for the non-local pairing interaction.

We can rewrite the Hamiltonian (\ref{Ham_K}) in the Gor$^{\prime}$kov-Nambu (or Bogoliubov-de Genes, BdG) representation, as is done for the spin-orbit coupled systems:
\begin{eqnarray}
\label{Ham_K_BdG}
\tilde{H} & = & \frac{1}{2} \sum_{k} \tilde{\Psi}^{\dag}_k \tilde{\mathcal{H}}_k \tilde{\Psi}_k,
\\
\tilde{\mathcal{H}}_k & = & \tau_z \otimes \left( \xi_k^{+} \sigma_0 - \sgn(h) t_k^- \sigma_x \right) - |h|\tau_0 \otimes \sigma_z +
\nonumber \\
& + & \tau_x \otimes \left[ \Re\left( \Delta_k^+ \right) \sigma_0 - \sgn(h) \Re\left( \Delta_k^- \right) \sigma_x \right] -
\nonumber \\
& - & \tau_y \otimes \left[ \Im\left( \Delta_k^+ \right) \sigma_0 - \sgn(h) \Im\left( \Delta_k^- \right) \sigma_x \right],
\end{eqnarray}
where $\tau_i$ and $\sigma_i$ are the Pauli matrices ($\tau_0$ and $\sigma_0$ are the $2 \times 2$ unit matrices of the same form) in the particle-hole space and spin space, respectively, $\otimes$ is the Kronecker product (further, we use the notation $\tau_i \otimes \sigma_j \equiv \tau_i \sigma_j$), the subscripts $\Re$ and $\Im$ denote the real and imaginary parts, respectively, and a spinor is introduced
\begin{equation}
\tilde{\Psi}^{\dag}_k = \left( c^{\dag}_{k \uparrow}, c^{\dag}_{k \downarrow}, c_{-k \downarrow}, -c_{-k \uparrow} \right).
\end{equation}
From now on, we will consider the nearest neighbor approximation for hoppings and pairings.

Finally, the Hamiltonian (\ref{Ham_K}) includes the Zeeman field along the $z$ axis and effective spin-flip processes (the second sum in (\ref{Ham_K})) with the function $t_k^{-}$ which is odd on $k$. For the square lattice, the function $t_k^{-}$ is
\begin{equation}
t_{s,k}^{-} = 2t_1\left[ \sin(k_x) \sin(Q_x/2) + \sin(k_y) \sin(Q_y/2) \right],
\end{equation}
and the index $s$ denotes the square lattice case.
Under the additional rotation around the $z$-axis at the angle $\pi/4$ and in the long-wave limit for $Q_x = Q_y = Q$ the following term is obtained in the BdG representation:
\begin{equation}
\label{Hso}
\tilde{\mathcal{H}}_{so}({\bf k}) = \sqrt{2}t_1 \sin(Q/2) \left( k_x  + k_y \right)\left( \sigma_x - \sigma_y \right),
\end{equation}
where $\sigma_i$ are the Pauli matrices in the spin space.
It can be seen \cite{martin-12} that this term is equivalent to the equal mixing of the Rashba spin-orbit interaction and 2D Dresselhaus term \cite{rashba-04}. Then, we can conclude that the helical spin ordering on the 2D lattice corresponds, as in the 1D case, to the uniform magnetic field in the presence of the effective spin-orbit interaction (\ref{Hso}). The difference between 1D and 2D is that in the 1D case the exact Rashba SOC is derived \cite{braunecker-10} while in the 2D case the corresponding term consists of more complex contributions. It is seen that for the ferromagnetic ordering ($Q = 0$) the term (\ref{Hso}) is absent.

Another crucial point for TSCty is the formation of effective triplet superconductivity. The presence of the effective triplet pairings in the Hamiltonian (\ref{Ham_K}) with the amplitude $\Delta_k^-$ is not so important; however, the mixing of fermions with the opposite spin moments plays a certain role. Indeed, there is a regime when all the processes involve fermions with the same spin. One of the ways to demonstrate it in the model (\ref{Ham_K}) is the analysis of band splitting due to the magnetic order \cite{martin-12}. Indeed, we can diagonalize the part of the Hamiltonian (\ref{Ham_K}) describing the helical spin ordering by the Bogoliubov transformation. Thus, the quasiparticle energy spectrum of the bands (the lower and upper bands are further referred to as d- and p-bands, respectively) is determined by
\begin{equation}
\label{epsk_mp}
\varepsilon_{k}^{\mp} = -\mu + t_k^{+} \mp \sgn(h)\sqrt{(t_k^{-})^2 + h({\bf Q})^2}.
\end{equation}
The hybridization of the bands is characterized by anomalous pairings between the states of different bands:
\begin{equation}
\sum_k \left(  A_k d_k^{\dag}p_{-k}^{\dag} + H.c. \right), \, \, \, A_k = \Delta^{+}_k h/\sqrt{(t_k^{-})^2 + h^2}.
\end{equation}

For the square lattice, if $|h|>|t^{s}_{Q/2}|$ there is an indirect gap between two bands (the bottom of the p-band lies above the top of the d-band). For $\mu < |t^{s}_{Q/2}| - |h|$ the d-band is only filled. Therefore, if $\text{max}\left( |\Delta_k^+| \right) \ll |h|$, when the processes between the bands can be neglected, the system is described by the d-band with the superconducting pairings between the states inside the band causing triplet superconductivity
\begin{eqnarray}
\label{Vk}
&&\sum_k \left( \frac{V^{-}_k}{2} d_k^{\dag}d_{-k}^{\dag} + H.c \right),
\nonumber \\
&& V^{\mp}_k = \sgn(h)\Delta_k^- \mp \Delta_k^+ t_k^{-}/\sqrt{(t_k^{-})^2 + h^2}.
\end{eqnarray}
It is seen that even for the local on-site pairing when $\Delta_k^+ = \Delta$, $\Delta_k^- = 0$ the effective triplet superconductivity can appear as a combination of the spin-singlet superconducting order parameter and amplitude $t_k^{-}$ (it is odd on $k$) of spin-flip processes. For $\mu > -|t^{s}_{Q/2}| + |h|$ the same considerations for the p-band can be given and the amplitude of pairings inside the p-band is $-V_k^+$.

\begin{figure*}
\center
\includegraphics[width=0.4\textwidth]{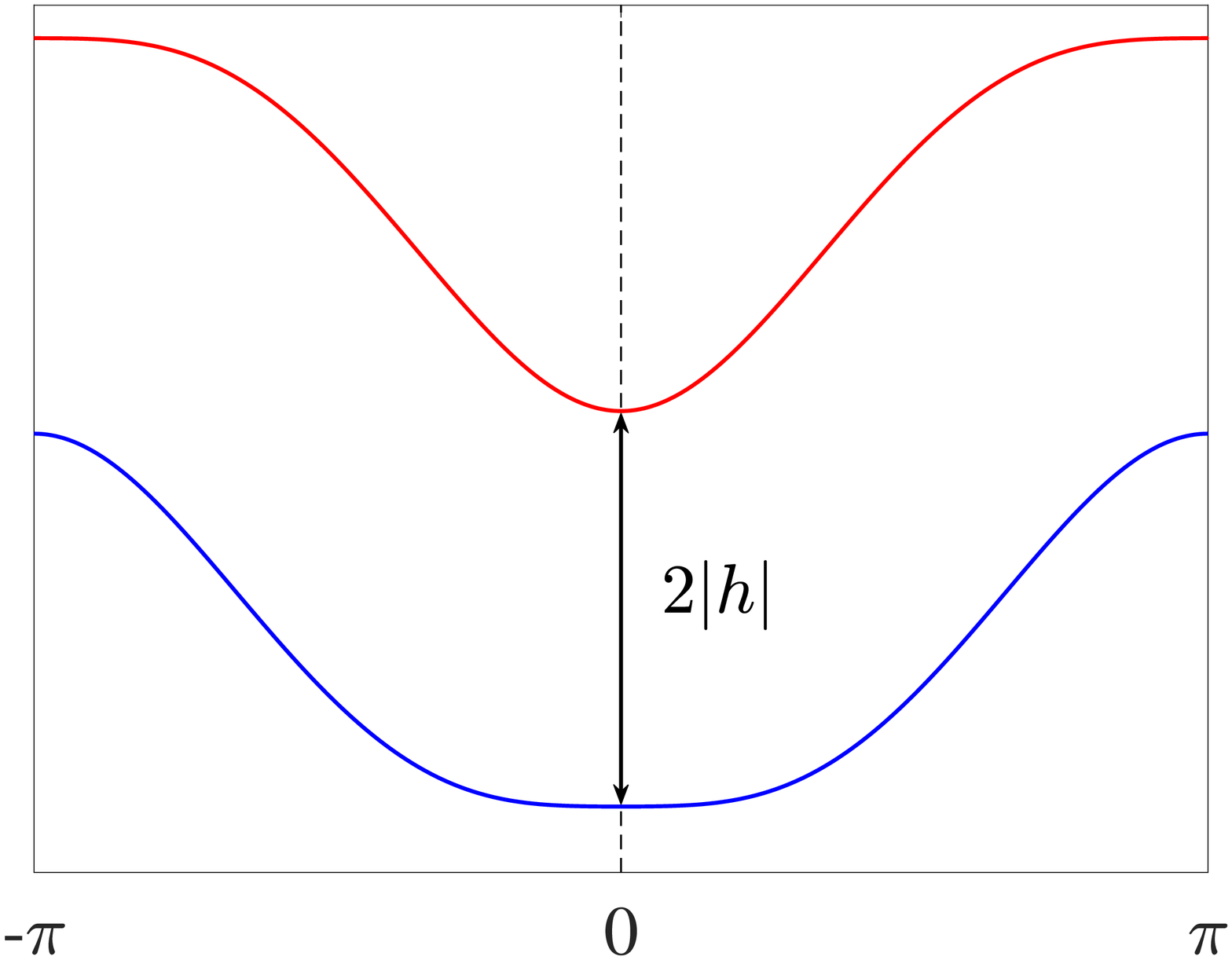}
\hfill
\includegraphics[width=0.4\textwidth]{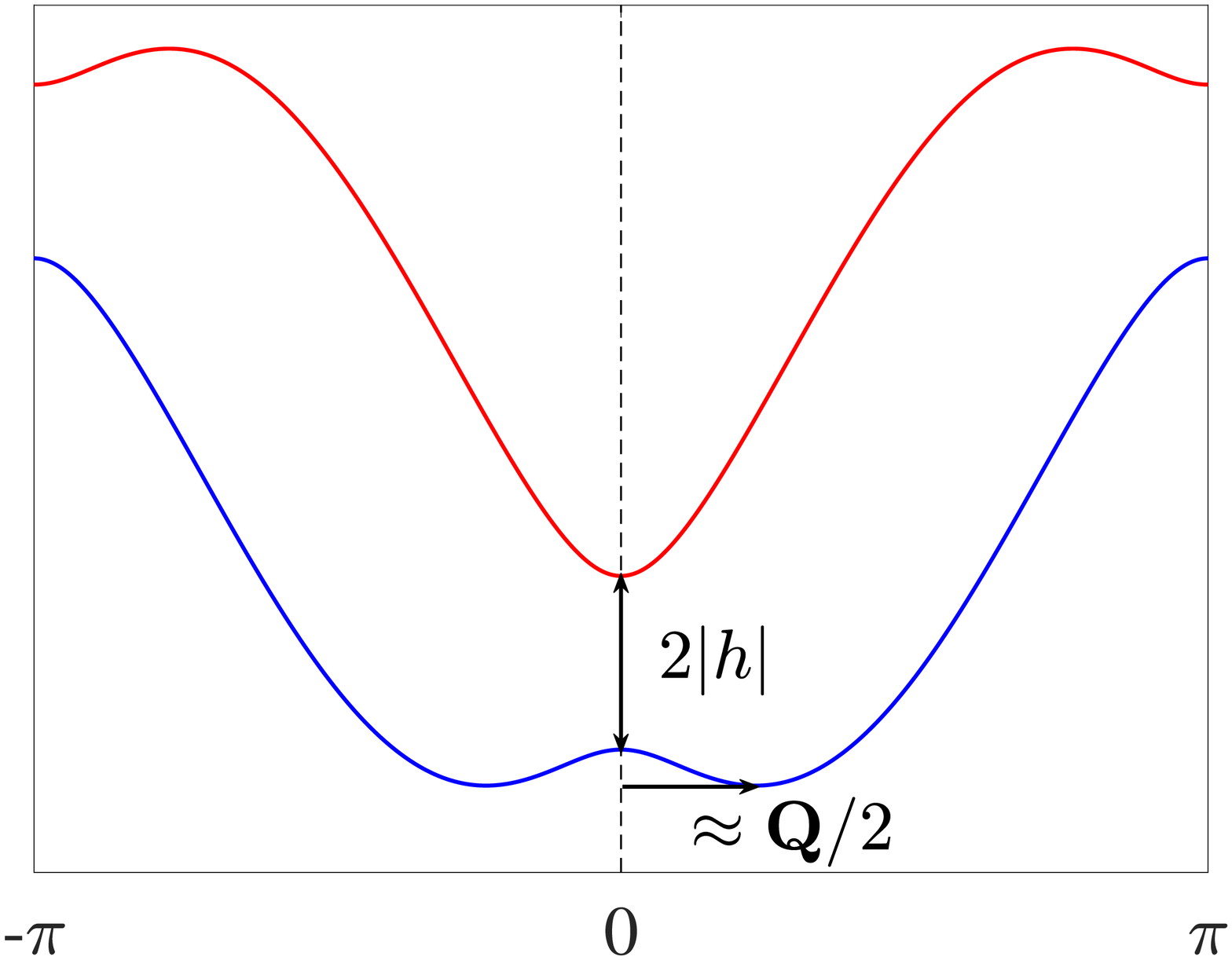}
\caption{The energy spectrum (\ref{epsk_mp}) in the main direction of the Brillouin zone in the presence of the noncollinear spin ordering with the wave-vector $\bf{Q}$ (see (\ref{Sf})) for the square lattice at $|h|>|t^s_{Q/2}|$ (the left plot)
and $|h|<|t^s_{Q/2}|$ (the right plot).}
\label{fig_bands}
\end{figure*}

If $|h|<|t^s_{Q/2}|$ the bands overlap, but for $\mu < -|t^s_{Q/2}| + |h|$ and $\mu > |t^s_{Q/2}| - |h|$, one-band filling is still realized and the effective triplet superconductivity can be induced for $\text{max}\left( |\Delta_k^{+}| \right) \ll |h|$. While for $-|t^s_{Q/2}| + |h| < \mu < |t^s_{Q/2}| - |h|$ the two band structure is essential. The two cases of the band structure for $|h|>|t^s_{Q/2}|$ and $|h|<|t^s_{Q/2}|$ are shown in Fig. \ref{fig_bands}.

It can be seen from expression (\ref{Vk}) for the triplet pairing terms that in the case of the ferromagnetic ordering with $Q=0$ the effective triplet superconductivity induced from the spin-singlet channel, which is described by the second term in (\ref{Vk}), is not realized since $t_k^- = 0$. Moreover, it is known that superconductivity is strongly destroyed by ferromagnetism while for antiferromagnetic and spiral structures it is only suppressed (see \cite{ashkenazi-83, buzdin-86,inui-88,kyung-00,lichtenstein-00,sacramento-03,demler-04,capone-06,weber-06,pathak-09,kuboki-10,sahoo-10,kaczmarczyk-11,VVV-ZAO-13, yamase-16, VVV-ZAO-16, foley-19, koshelev-19} and references therein).

It is seen from definition (\ref{Vk}) that the first term of the expansion of $V_k$ in the long-wave limit and for $\left| h \right| \gg |t_1|$ is linear in ${\bf k}$. Firstly, the realization of  this $p_x+p_y$ superconducting state due to the spin helix state in the 2D system with the initial s-wave superconductivity was demonstrated in \cite{nakosai-13, chen-15}. In \cite{nakosai-13} the unitary transformation similar to (\ref{U}) for the general direction of the spin moments (\ref{GSf}) is performed and the effective parameters of hoppings and superconducting pairings are calculated. It should be noted that the $p_x+p_y$ superconducting gap is a nodal gap with zeros on the specific lines in the Brillouin zone. It is worth noting that in the same approach the noncoplanar spin texture (for example, the skyrmion state, see the details in the next section) induces the fully gapped chiral $p_x+ip_y$ superconductivity which is more suitable to search for Majorana fermions.

For the triangular lattice, the function $t_k^{-}$ is
\begin{eqnarray}
t_{t,k}^{-} & = & 2t_1\left[ \sin(k_1) \sin(Q_1/2) + \sin(k_2) \sin(Q_2/2) + \right.
\nonumber \\
& + & \left. \sin(k_1+k_2) \sin(Q_1/2+Q_2/2) \right],
\end{eqnarray}
where the components $k_i$, $Q_i$ of the wave-vectors are chosen on the basis of the elementary vectors ${\bf b_i}$ of a reciprocal lattice, such as ${\bf k} = k_1 {\bf b_1} + k_2 {\bf b_2}$. We also use the definition of the radius-vector in the real space ${\bf R}_f = f_1 {\bf a_1} + f_2 {\bf a_2}$, where $f_i$ are the integer numbers, and the ``crystallographer'' notation ${\bf a_i} {\bf b_j} = \delta_{ij}$. Again, $Q_i \in (0, \pi)$. In this case, the indirect gap between the bands is formed if
\begin{equation}
|h| > g = 2 |t_1| \left( \cos(Q_1/2) + \cos(Q_2/2) \right).
\end{equation}

The conditions for one-band filling on the triangular lattice are as follows:
\begin{eqnarray}
\label{cond_triplet}
\text{for} \, \, \, |h| < g \, \, \, \varepsilon_{\text{bot}}^- < \mu < \varepsilon_{\text{bot}}^+ \, \, \, \text{and} \, \, \, \varepsilon_{\text{top}}^-<\mu<\varepsilon_{\text{top}}^+,
\nonumber \\
\text{for} \, \, \, |h| > g \, \, \, \varepsilon_{\text{bot}}^- < \mu < \varepsilon_{\text{top}}^- \, \, \, \text{and} \, \, \, \varepsilon_{\text{bot}}^+<\mu<\varepsilon_{\text{top}}^+.
\end{eqnarray}
The energies at the top of the d-band and at the top and bottom of the p-band  (\ref{epsk_mp}) for the triangular lattice are determined by ($t_1 < 0$)
\begin{eqnarray}
\varepsilon_{\text{top}}^- & = & 2|t_1|\left( \cos(Q_1/2) + \cos(Q_2/2) \right) +
 \nonumber \\
& + & 2 t_1 \cos((Q_1+Q_2)/2) - |h|,
\\
\varepsilon_{\text{bot}}^+ & = & -2|t_1|\left( \cos(Q_1/2) + \cos(Q_2/2) \right) +
\nonumber \\
& + & 2t_1 \cos((Q_1+Q_2)/2) + |h|,
\\
\label{eps_top_p}
\varepsilon_{\text{top}}^+ & = & -2t_1\left( \cos(Q_1/2) + \cos(Q_2/2) \right) +
\nonumber \\
& + & 2t_1 \cos((Q_1+Q_2)/2) - \sgn(t_1) |h|.
\end{eqnarray}
In general, the bottom of the d-band is located in the vicinity of the ${\bf Q}/2$ point in BZ. However, to find the exact value of $\varepsilon_{\text{bot}}^-$ a system of equations should be solved. For the 120$^{\circ}$ ordering with $Q_1=Q_2=2\pi/3$ the bottom of the d-band is located exactly at ${\bf Q}/2$ for any model parameter values. Therefore, the expression for $\varepsilon_{\text{bot}}^-$ can be written as
\begin{equation}
\label{eps_bot_m}
\varepsilon_{\text{bot}}^- = 3t_1 + t_{t, Q}/2 + \sgn(t_1) \sqrt{(3t_1 - t_{t, Q}/2)^2+h^2}.
\end{equation}
For $t_1 > 0$ the expressions for $\varepsilon_{\text{top}}^-$, $\varepsilon_{\text{bot}}^+$ remain the same, while the energy (\ref{eps_bot_m}) corresponds to the top of the p-band, otherwise the energy (\ref{eps_top_p}) corresponds to the bottom of the d-band.

The above analysis relates only to filling of the d- and p-bands and the obtained conditions (\ref{cond_triplet}) are a rough estimate of the triplet superconductivity regime. A more rigorous result can be obtained by the Schrieffer-Wolff transformation method. It shows that the triplet superconductivity is implemented in the regime of a strong exchange field with $|h| \gg |t_1|, \, |\Delta|$ and $|\mu| \approx \left| h \right|$. In this case normal and anomalous processes involving states from different spin bands can be taken into account in the perturbation theory up to the terms $\sim t_1^2/|h|$, $\sim |t_1 \Delta/h|$, and $\sim |\Delta^2/\mu|$ where $\Delta$ is the amplitude of the superconducting order parameter. In the 1D case, such a transformation was demonstrated in \cite{choy-11}. There is the straightforward generalization of this transformation to the 2D case \cite{chen-15}. It should be noted that the triplet pairing term $\sim |t_1 \Delta/h|$ can easily be obtained as the first term of the expansion of $V_k^{\mp}$ (\ref{Vk}) in the considered limit.

\subsection{Topological invariants and Majorana fermions in 2D magnetic superconductors}

Different topological invariants to describe a wide range of topological superconductors are proposed. In the following, we will briefly discuss topological invariants for magnetic superconductors.

First of all, it should be noted that if $\Delta_k^{\pm}$ are the real functions, then the BdG Hamiltonian (\ref{Ham_K_BdG}) can easily be transformed to an off-diagonal view:
\begin{equation}
W^{\dag} \tilde{\mathcal{H}}_k W = \left(
                                         \begin{array}{cc}
                                           O & D_k \\
                                           D_k^{\dag} & O \\
                                         \end{array}
                                       \right),
\, \, \, W = \left( \tau_z \sigma_0 - \tau_y\sigma_x \right).
\end{equation}
Therefore, the topological invariant can be determined in the spirit of spin-orbit coupled superconductor nano\-wires \cite{tewari-12} as the winding number which is characterized by $\text{Det}\left( D_k \right)$. This invariant can be calculated through the particle-hole invariant momenta (PHIMs) which are $\tilde{\textbf{k}} = -\tilde{\textbf{k}} + \textbf{G}$ (${\bf G}$ is a reciprocal lattice vector) for the transformed Hamiltonian. The details of this approach for TSCs with magnetic textures are presented in \cite{steffensen-20}. Nevertheless, for the 2D lattice, this topological invariant is calculated only for the fixed quasi-momentum $k_2$ in the second direction of the reciprocal space.

It is supposed that the consideration of the 2D system, with one of two quasi-momenta (e.g. $k_2$) being fixed, is justified when it has the form of a long stripe. Then, in the center of the stripe we consider only the edges in the transverse direction while the effects of the other two edges are negligible. Therefore, the periodic boundary conditions can be applied in the direction along the stripe. In conclusion, the 2D system can be divided into a set of independent 1D systems having the appropriate quantum number $k_2$. For the fixed $k_2$, it is possible to determine the 1D topological invariant called the Majorana number \cite{kitaev-01} and to obtain conditions for the formation of the Majorana edge states in the transverse direction \cite{chen-15, VVV-ZAO-ShMS-18}.

For the 1D case, it is believed that the topological invariant (e.g. Majorana number \cite{kitaev-01}) is connected with the fermion parity of the ground state, for example, for the non-trivial topological phase the wave function of the ground state is determined by the superposition of the states with odd fermion numbers (further referred to as the odd function). As is known, the Bardeen - Cooper - Schrieffer wave function is even; therefore, conventional superconductors are trivial and there are no pronounced effects due to edges or defects. Adjusting the Bardeen - Cooper - Schrieffer superconductor in such a way that the ground state wave function becomes odd, opening the way for TSCty and Majorana fermions \cite{read-00}.

For the quadratic Hamiltonian, the parity of the ground state is determined only by filling of the states in specific points of the Brillouin zone. Consider the 2D triangular lattice with the 120$^{\circ}$  spin ordering and chiral d-wave superconductivity \cite{VVV-ZAO-FAD-ShMS-17, VVV-ZAO-ShMS-18} being described by Hamiltonian (\ref{Ham}). For this case, the indices $f$ and $m$ in Hamiltonian (\ref{Ham}) label the sites of the triangular lattice and the wave-vector of the 120$^{\circ}$ spin structure is ${\bf Q} = \left( 2\pi/3, 2\pi/3\right)$. It is known \cite{baskaran-03, zhou-08, VVV-VTA-MVA-15} that the symmetry of the triangular lattice supports the formation of the chiral superconducting order parameter if the orbital number characterizing the symmetry type of superconductivity is $l \ne 0$. We suppose the formation of an energetically favorable spin-singlet chiral $d_1+id_2$ wave superconducting state with $l = 2$.

In the following, for simplicity we will use the initial Hamiltonian (\ref{Ham}) instead of the transformed Hamiltonian (\ref{Ham_K}). With the periodic boundary condition, Hamiltonian (\ref{Ham}) in the momentum space can be written as
\begin{eqnarray}
\label{Ham_K2}
&& H = \frac{1}{2} \sum_{k} \Psi^{\dag}_k \mathcal{H}_k \Psi_k,
\, \, \,
\Psi^{\dag}_k = \left( c^{\dag}_{k \uparrow}, c^{\dag}_{k-Q \downarrow}, c_{-k+Q \uparrow}, c_{-k \downarrow} \right),
\nonumber \\
&& \mathcal{H}_k = \left(
                                         \begin{array}{cc}
                                           A(k) & B(k) \\
                                           -B^*(-k+Q) & -A^*(-k+Q) \\
                                         \end{array}
                                       \right),
\\
&& A(k) = \left(
                                         \begin{array}{cc}
                                           \xi_k & h \\
                                           h & \xi_{k-Q} \\
                                         \end{array}
                                       \right), \, \, \,
B(k) = \left(
                                         \begin{array}{cc}
                                           0 & \Delta_k \\
                                    \Delta_{-k+Q} & 0 \\
                                         \end{array}
                                       \right),
\nonumber
\end{eqnarray}
where for the triangular lattice $\xi_k = -\mu + t_{t, k}$ and
\begin{eqnarray}
\Delta_{t, k} & = & 2\Delta_{21}\left[ \cos(k_1) + e^{i2\pi/3} \cos(k_1+k_2) + \right.
\nonumber \\
& + & \left. e^{i4\pi/3} \cos(k_2) \right].
\end{eqnarray}
The bulk energy spectra of Hamiltonians (\ref{Ham_K2}) and (\ref{Ham_K}) are connected by the replacement $\tilde{{\bf k}} \to {\bf k}-{\bf Q}/2$ and vice versa.

Hamiltonian (\ref{Ham_K2}) supports only the particle-hole symmetry and belongs to the D class. Therefore, the Hamiltonian has the property:
\begin{eqnarray}
\Lambda \mathcal{H}_k \Lambda = - \mathcal{H}^*_{-k+Q}, \, \, \, \Lambda = \left(
                                         \begin{array}{cc}
                                           O & I \\
                                           I & O \\
                                         \end{array}
                                       \right).
\end{eqnarray}
$O$ and $I$ are zero and unit 2 by 2 matrices respectively. A direct consequence of the symmetry is that the excitation energy $\varepsilon_{-k + Q}$ is the same as the energy $\varepsilon_{k}$. Therefore, for ${\bf k} \ne -{\bf k}+{\bf Q}+{\bf G}$ the fermion states are filled only by pairs and the parity of the ground state is not changed due to this filling. To obtain conditions of the parity change the filling of the fermion states corresponding to PHIMs in the presence of the noncollinear spin ordering ${\bf K} = -{\bf K}+{\bf Q}+{\bf G}$ should be analyzed. It should be noted that the same analysis for the topological systems with the spin-orbit interaction (${\bf Q}=(0,0)$) was carried out \cite{VVV-MVA-ShMS-19}.
For the triangular lattice with the 120$^{\circ}$ ordering there are four PHIMs ${\bf K}_1 = (-2\pi/3,-2\pi/3)$, ${\bf K}_2 = (\pi/3,\pi/3)$, ${\bf K}_3 = (-2\pi/3,\pi/3)$, ${\bf K}_4 = (\pi/3,-2\pi/3)$.

\begin{figure*}
\center
  \includegraphics[width=0.6\textwidth]{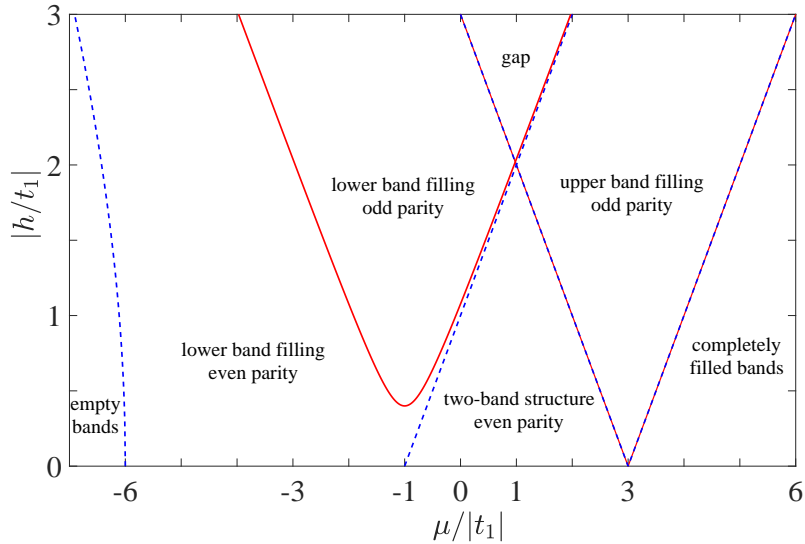}
\caption{Conditions for the one-band filling and even or odd fermion parity of the ground state with the coexisting superconductivity and 120$^{\circ}$ spin ordering on the triangular lattice. The variables are chemical potential ($\mu$) and exchange field ($h$), $t_1$ is the hopping parameter between the nearest sites.}
\label{fig_fill_par}
\end{figure*}

Let us analyze Hamiltonian (\ref{Ham_K2}) with PHIMs. For the point ${\bf K}_1$ the superconducting order parameter $\Delta_k$ is zero, then at this point the Hamiltonian has the form:
\begin{eqnarray}
H_{K_1} & = & (-\mu-3t_1) \left( c^{\dag}_{K_1 \uparrow} c_{K_1 \uparrow} + c^{\dag}_{-K_1 \downarrow} c_{-K_1 \downarrow} \right) +
\nonumber \\
& + & h \left( c^{\dag}_{K_1 \uparrow} c_{-K_1 \downarrow} + c^{\dag}_{-K_1 \downarrow} c_{K_1 \uparrow} \right).
\end{eqnarray}
The corresponding energies of Bogoliubov quasi-particles are
\begin{equation}
\varepsilon^{\mp}_{1} = -3t_1 - \mu \mp |h|.
\end{equation}
The same procedure for the point ${\bf K}_2$ can be done:
\begin{eqnarray}
&& H_{K_2} = (-\mu + t_1) \left( c^{\dag}_{K_2 \uparrow} c_{K_2 \uparrow} + c^{\dag}_{-K_2 \downarrow} c_{-K_2 \downarrow} \right) +
\nonumber \\
& + &  h \left( c^{\dag}_{K_2 \uparrow} c_{-K_2 \downarrow} + c^{\dag}_{-K_2 \downarrow} c_{K_2 \uparrow} \right) +
\nonumber \\
& + & \left( \Delta_{21}(1-\sqrt{3})c^{\dag}_{K_2 \uparrow} c^{\dag}_{-K_2 \downarrow} + \Delta_{21}^*(1+\sqrt{3})c_{-K_2 \downarrow} c_{-K_2 \uparrow} \right).
\nonumber \\
\end{eqnarray}
In this case, the Bogoliubov quasi-particle energies are
\begin{equation}
\varepsilon^{\mp}_{2} = \sqrt{(t_1 - \mu)^2 +4|\Delta_{21}|^2} \mp |h|.
\end{equation}
The quasi-particles for the remaining PHIMs ${\bf K}_3$ and ${\bf K}_4$ have the same energy $\varepsilon^{\mp}_{2}$. If the odd number of the obtained quasi-particle states is filled ($\varepsilon<0$), then the odd-parity of the ground state function of the whole system is realized. It can easily be shown that the conditions for the odd-parity ground state are
\begin{eqnarray}
\label{cond_odd}
|-3t_1 - \mu|<|h|<\sqrt{(t_1 - \mu)^2 +4|\Delta_{21}|^2},
\end{eqnarray}
or depending on the considered interval of the chemical potential
\begin{eqnarray}
\label{cond_odd2}
\sqrt{(t_1 - \mu)^2 +4|\Delta_{21}|^2}<|h|<|-3t_1 - \mu|.
\end{eqnarray}
To change the parity of the ground state the gap of the bulk spectrum must become zero with the odd number of points in BZ. For example, the gap is closed at PHIM ${\bf K}_1$ for $|h| = |-3t_1 - \mu|$ and at PHIMs ${\bf K}_1$, ${\bf K}_2$, ${\bf K}_3$ for $|h| = \sqrt{(t_1 - \mu)^2 +4|\Delta_{21}|^2}$, leading to the change in the parity. When the spectrum becomes gapless at the even number of points in BZ the parity does not change.

In Fig. \ref{fig_fill_par} the regions corresponding to conditions (\ref{cond_odd}) and (\ref{cond_odd2})  with the odd parity of the ground state are shown by the solid lines on the diagram in the variables of the chemical potential $\mu$ and exchange field (to be specific, we choose $t_1 < 0$ and $\Delta_{21} = 0.2 |t_1|$). By the dashed lines we also denote qualitative conditions (\ref{cond_triplet}) for the one-band filling when the pairings between the d- and p-bands can be neglected (for $\Delta_{21} \ll |h|$) and the pairings inside one band become suitable. It is supposed that the initial spin-singlet superconducting pairing is formed on the whole diagram. This case can be achieved in magnet - superconductor hybrids, when superconductivity is proximity-induced. On the other hand, for magnetic superconductors the superconducting region should be determined from self-consistent equations. To consider the effective one-band filling in certain regions of the diagram the limit $\Delta_{21} \ll |h|$ should be considered. Therefore, in the region with $|h| \ll |t_1|$ the superconducting gap vanishes and this regime is of no practical interest.

It is seen from Fig. \ref{fig_fill_par} that the odd-parity conditions are satisfied only inside the regions with one-band filling when effective triplet pairings can be formed. Therefore, inequalities (\ref{cond_triplet}) are the necessary, but not sufficient, conditions for the topologically nontrivial phases. Otherwise, as it will be discussed later, the conditions (\ref{fig_fill_par}) are sufficient to describe the topologically nontrivial phases with the Majorana fermions.

As is known for the antiferromagnetic ordering $Q_i = \pi$, there is always a gap between the bands (\ref{epsk_mp}) for $h \ne 0$. Moreover, the regions with the odd fermion parity and topologically nontrivial phases collapse to zero for the antiferromagnetic ordering. Therefore, this magnetic structure does not support MMs.

It is known that the effective time-reversal symmetry can be present in the exact 1D theoretical models similar to (\ref{Ham_K_BdG}) even in the presence of the magnetic field (see for example \cite{tewari-12, wong-12, poyhonen-14, sedlmayr-15, ASV-ZAO-ShMS-20, steffensen-20}). Such an additional symmetry leads to the BDI class topological symmetry instead of the D class. For the BDI class symmetry, the time-reversal, particle-hole and chiral symmetries are present (or their analogs), while there is only particle-hole symmetry for the D class. Usually, the differences in the topological features between two classes appear only when the long-range or even next-nearest electron hoppings and pairings are taken into account. Nevertheless, for the 2D system (\ref{Ham_K_BdG}) the D class symmetry is believed to be implemented \cite{VVV-ZAO-ShMS-18, bedow-20}. In the following, we will also focus on the chiral d-wave superconductivity breaking the time-reversal symmetry even in the absence of the magnetic field.

For the D class symmetry, in 2D the topological invariant is the Chern number given by
\begin{equation}
C = -\frac{1}{2\pi} \iint\limits_{-\pi}\limits^{\pi} dk_1dk_2 F_{12} ({\bf k})
\end{equation}
and $F_{12} = \partial_1 A_2({\bf k}) - \partial_2 A_1({\bf k})$ is the Berry curvature, $A_{\mu}({\bf k}) = -i \left \langle n({\bf k}) \right| \partial_{\mu} \left| n({\bf k}) \right \rangle$ is the $\mu$-component of the Berry vector potential (or Berry connection),  $\partial_{1} \equiv \partial/\partial k_1$,
$\partial_{2} \equiv \partial/\partial k_2$, $\left| n({\bf k}) \right \rangle$ is the Bloch state (for simplicity, we assume the one-band structure). The examples of the calculation of the topological phase diagrams of magnetic superconductors through the Chern number can be found in \cite{crawford-20, bedow-20}.

The generalization of the Chern number via the Green functions is also known ~\cite{ishikawa-87, volovik-03}:
\begin{eqnarray}
\label{Z_inv} \tilde{N}_{3}& = &
\frac{\varepsilon_{\mu\nu\lambda}}{24\pi^{2}} \times
\\
&\times&
\int\limits_{-\infty}^{\infty}d\omega\iint\limits_{-\pi}\limits^{\pi}
dk_{1}dk_{2}
\text{Tr}\left(\widehat{G}\partial_{\mu}\widehat{G}^{-1}
\widehat{G}\partial_{\nu}\widehat{G}^{-1}
\widehat{G}\partial_{\lambda}\widehat{G}^{-1} \right). \nonumber
\end{eqnarray}
Here, the repeated indices $\mu$, $\nu$, $\lambda = 1, \, 2, \, 3$
imply summation,  $\varepsilon_{\mu\nu\lambda}$ is the Levi-Civita symbol, $\partial_{3} \equiv
\partial/\partial \omega$, and $\widehat{G}(i\omega,k)$ is the matrix
Green function whose poles determine the spectrum
of elementary fermion excitations. The integral on $\omega$ can be taken analytically since the excitation spectrum is gapped. It should be noted that the topological invariant $\tilde{N}_{3}$  can also be used in interacting systems. The details of the calculation of $\tilde{N}_{3}$ in the chiral $d_1+id_2$ superconductor with the 120$^{\circ}$ spin order in the presence of strong electron correlations on the triangular lattice are presented in \cite{VVV-ZAO-19}.

\begin{figure}
\begin{center}
  \includegraphics[width=0.45\textwidth]{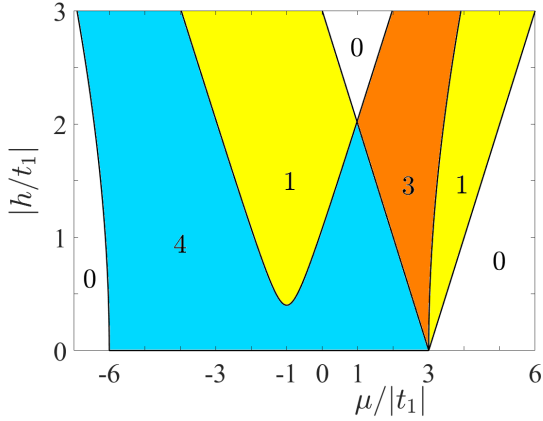}
\caption{Topological phase diagram in the variables of chemical potential ($\mu$) and exchange field ($h$) (see also Fig. \ref{fig_fill_par}). Different colors denote different topological phases which are determined by the topological invariant $\tilde{N}_3$ (see (\ref{Z_inv})). The values of $\tilde{N}_3$ are marked. The topologically trivial phases are the phases with $\tilde{N}_3 = 0$, the nontrivial phases have $\tilde{N}_3 \ne 0$. The Majorana fermions are realized in the phases with odd $\tilde{N}_3$.}
\label{fig_top_lines}
\end{center}
\end{figure}

The topological phase diagram calculated through the topological invariant $\tilde{N}_{3}$ is presented in Fig. \ref{fig_top_lines} in the same variables and limits as in Fig. \ref{fig_fill_par}. First of all, we should note that even in the absence of the long-range magnetic order ($h = 0$) there is a topologically nontrivial phase with $\tilde{N}_{3} = 4$ corresponding to the case of the chiral $d_1 + id_2$ superconductor \cite{volovik-97, zhou-08, VVV-VTA-MVA-15}. It is known that topologically protected edge states exist in such a superconductor. Nevertheless, the Majorana end states are prohibited at $h = 0$ due to the spin-singlet character of superconducting pairings.

It is seen from Fig. \ref{fig_top_lines} that the formation of the long-range spin ordering with a sufficiently strong exchange field leads to topological phase transitions changing the parity of  $\tilde{N}_{3}$.
It is believed that the odd values of the topological invariant $\tilde{N}_{3}$ indicate the non-abelian topological order supporting the Majorana fermions. This statement for noncentrosymmetric superconductors with the broken time-reversal symmetry was derived in \cite{ghosh-10}. In general, the same arguments are satisfied for magnetic superconductors with the broken time-reversal symmetry bearing in mind that a new set of PHIMs is  ${\bf K} = -{\bf K}+{\bf Q}+{\bf G}$. Therefore, the increasing exchange field modifies the initial end states of the chiral superconductor to the Majorana end states.

As can be seen from Fig. \ref{fig_top_lines} the topological phases with odd $\tilde{N}_{3}$ exactly coincide with the conditions of odd parity of the ground state from Fig. \ref{fig_fill_par}. At each topological phase transition the bulk spectrum of $\mathcal{H}_k$ (\ref{Ham_K2}) becomes gapless. It is seen that there is the transition without changing the fermion parity (from the phase with $\tilde{N}_{3} = 3$ to the phase with $\tilde{N}_{3} = 1$). It is necessary to taking into account that the superconducting pairings between the electrons on the next nearest sites lead to the appearance of new topological phase transitions \cite{VVV-ZAO-ShMS-18}. Nevertheless, these transitions do not change the fermion parity and the conditions for the MBS existence remain the same.

To reveal the appearance of MBS in such phases it is convenient to consider a long strip instead of a 2D lattice when the periodic boundary conditions along the strip can be applied. In this case, the quantum number, for example $k_2$, is well defined and the system can be studied independently at different $k_2$.

The BdG Hamiltonian $\mathcal{H}^{\text{strip}}_{k_2}$ at the fix $k_2$ having the size $4N_1 \times 4N_1$ ($N_1$ is the strip size in the transverse direction)  can be determined as
\begin{equation}
\label{Ham_K2_strip}
H^{\text{strip}}_{k_2} = [\hat{c}^{\dag} \hat{c}^{T}] \mathcal{H}^{\text{strip}}_{k_2}  \left[
                                         \begin{array}{c}
                      \hat{c}\\
                     \hat{c}^{*}              \end{array} \right],
\end{equation}
where $\hat{c} = \left[\hat{c}_{\uparrow} \hat{c}_{\downarrow} \right]^{T}$ and $\hat{c}_{\sigma}$ is the vector consisting of the Fermi operators $c_{l \sigma}$ with the index $l$ of a site of the strip at the fixed $k_2$.
Preforming the unitary (or Bogoliubov) transformation $V$ the following Hamiltonian is obtained
\begin{equation}
\label{Ham_a}
H^{\text{strip}}_{k_2} =  [\hat{\alpha}^{\dag} \hat{\alpha}^T] \hat{E}_D(k_2) \left[
                                         \begin{array}{c}
                      \hat{\alpha}\\
                     \hat{\alpha}^{*}              \end{array} \right]
\end{equation}
and $\hat{\alpha}$ is the set of the $2N_1$ Bogoliubov quasiparticle operators, $\hat{E}_D(k_2)$ is the diagonal matrix determining positive and negative energies of $2N_1$. Due to the particle-hole symmetry the positive and negative energies constitute pairs with the same absolute values. We suppose that the first $2N_1$ elements of $\hat{E}_D(k_2)$ are positive in the ascending order while the last elements are negative. Therefore, the term of the Hamiltonian $H^{\text{strip}}_{k_2}$ corresponding to the minimum excitation energy is $\varepsilon_1 \alpha_1^{\dag} \alpha_1$. To find MBS, it is necessary that $\varepsilon_1 = 0$.

\begin{figure}
\begin{center}
  \includegraphics[width=0.45\textwidth]{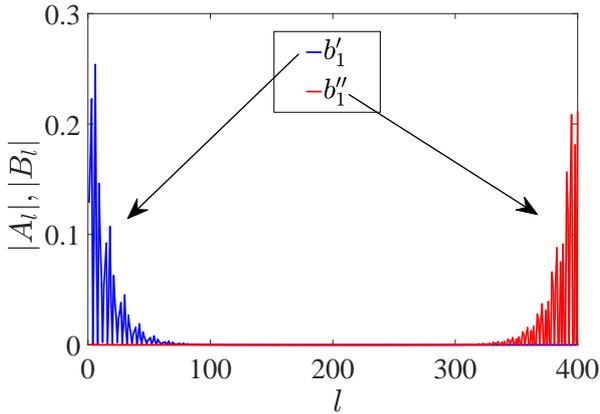}
\caption{Typical Majorana modes determined by (\ref{b_gamma}) in the quasi-1D case (the strip at $k_2 = -2\pi/3$) with the coexisting superconductivity and noncollinear spin ordering.}
\label{fig_maj_modes}
\end{center}
\end{figure}

The connection between the Bogoliubov and Fermi operators is written as
\begin{equation}
\left[
                                         \begin{array}{c}
                      \hat{\alpha}\\
                     \hat{\alpha}^{*}              \end{array} \right] = V^{\dag} \left[
                                         \begin{array}{c}
                      \hat{c}\\
                     \hat{c}^{*}              \end{array} \right].
\end{equation}
To define MMs, the Majorana representation \cite{kitaev-01} with the new Majorana operators is used
\begin{eqnarray}
\label{gammaA}
\gamma_{A l \sigma} & = & c_{l \sigma} + c_{l \sigma}^{\dag},
\\
\label{gammaB}
\gamma_{B l \sigma} & = & i\left(c_{l \sigma}^{\dag} - c_{l \sigma} \right).
\end{eqnarray}
Then
\begin{equation}
\label{alpha_gamma}
\left[
                                         \begin{array}{c}
                      \hat{\alpha}\\
                     \hat{\alpha}^{*}              \end{array} \right] = V^{\dag} S \left[
                                         \begin{array}{c}
                      \hat{\gamma}_{A}\\
                     \hat{\gamma}_{B}              \end{array} \right], \, \, \,
   S = \frac{1}{2} \left[                      \begin{array}{cccc}
                      \hat{I}_{N_1} & \hat{O}_{N_1} & i\hat{I}_{N_1} & \hat{O}_{N_1}\\
                     \hat{O}_{N_1} & \hat{I}_{N_1} & \hat{O}_{N_1} & i\hat{I}_{N_1}\\
                     \hat{I}_{N_1} & \hat{O}_{N_1} & -i\hat{I}_{N_1} & \hat{O}_{N_1}\\
                     \hat{O}_{N_1} & \hat{I}_{N_1} & \hat{O}_{N_1} & -i\hat{I}_{N_1}
                     \end{array} \right],
\end{equation}
and $\hat{I}_{N_1}$, $\hat{O}_{N_1}$ are the unit and zero matrices of the size $N_1 \times N_1$.

Topological arguments significantly simplify searching for the solution of (\ref{Ham_a}) corresponding to MMs. It is expected that the particle-hole symmetry protected MBS are formed for $k_2 = K_2$, where $K_2$ is the corresponding component of PHIMs (for 120$^{\circ}$ ordering $K_2 = -2\pi/3, \, \pi/3$).
In \cite{VVV-ZAO-ShMS-18} we show that MBSs exist at $K_2 = -2\pi/3$ and the conditions of their realization are well-defined by the topological phases with the odd topological invariant $\tilde{N}_{3}$ and, consequently, odd fermion parity. Moreover, the calculation of the Majorana number \cite{kitaev-01} at $K_2 = -2\pi/3$ leads to the same topologically nontrivial phases as to those having the odd invariant $\tilde{N}_{3}$ in Fig. \ref{fig_top_lines}.

The spatial structure of MMs can be found as follows. Using the Bogoliubov operators $\alpha$ the Majorana quasiparticle operators are defined as
\begin{equation}
\label{b_gamma}
\left[
                                         \begin{array}{c}
                      \hat{b}^{\prime}\\
                     \hat{b}^{\prime \prime}              \end{array} \right] = \left[
                                         \begin{array}{c}
                      \hat{\alpha}+\hat{\alpha}^*\\
                     i\left( \hat{\alpha}^* - \hat{\alpha} \right)              \end{array} \right].  \end{equation}
From the relation connecting the Majorana quasiparticle operators $b_j^{\prime}$ and $b_j^{\prime \prime}$ with the Majorana operators $\gamma_{A l \sigma}$ and $\gamma_{B l \sigma}$ (it can be derived using (\ref{alpha_gamma})) the coefficients depending on the excitation energy and site $l$ are determined. In general, the representation (\ref{b_gamma}) can be used for each excitation energy, and not only for zero energy. Nevertheless, only the zero mode solution should be analyzed. This solution is Majorana if $b_1^{\prime}$ and $b_1^{\prime \prime}$ are spatially separated, for example, on different edges. Thus, $b_1^{\prime}$ and $b_1^{\prime \prime}$ are often considered as two Majorana modes. The typical MBS in the TSC with the triangular lattice and 120$^{\circ}$ spin ordering is shown in Fig. (\ref{fig_maj_modes}).

The same procedure can be also used to find Majorana fermions in the 2D lattice. Then, the BdG Hamiltonian (\ref{Ham_K2_strip})
has the size $4N \times 4N$ and $N = N_1N_2$. There are few problems for limited 2D systems. Firstly, the Majorana solution becomes localized on the whole 1D edge of the 2D lattice (this is not the case of higher-order topological superconductors, see Sec. 4). Thus, there is no spatial resolution of different MMs. Secondly, if a superconductor is nodal (like extended s-wave or d-wave), i.e. the gap is closed in the bulk spectrum at the nodal lines of BZ, then such trivial bulk solutions can have near zero energy and admix to the Majorana solution in the 2D case. The second problem is solved in chiral superconductors having predominantly a gapped spectrum. MBS in such superconductors are separated by an energy gap from other edge states having finite energy. It should be noted that this gap may be much smaller than the bulk gap. The first problem could be solved considering the Abrikosov vorticies and different defects which can be controlled. In the topologically nontrivial phases the Majorana modes are formed on such defects and there appears a possibility of their braiding. Among the mentioned defects one of the intensively studied objects is a skyrmion. Its noncoplanar spin structure causes TSCty similarly to the long-range spin ordering. However, unlike the uniform long-range order, the localization of Majorana modes depends on the structure of the skyrmion. The details are presented in the next section.

To conclude this section, we show that the effective triplet superconductivity is possible and the odd fermion parity of the ground state is realized in the coexistence phase of the spin-singlet superconductivity and noncollinear spin ordering. These features cause the formation of topologically nontrivial phases supporting the Majorana modes.

\section{Topological superconductivity in 2D magnetic skyrmions}
\label{sec:3}

\subsection{The main properties of skyrmions}
\label{sec:3.1}

With the start of the pioneering research \cite{bloch-30, landau-65} topological objects began to attract considerable attention in the physics of magnetism. F. Bloch was the first to make an analytical description of the domain wall in ferromagnets \cite{bloch-30}, while Landau and Lifshitz developed a more complete phenomenological theory \cite{landau-65}. Currently, topological defects in magnetism include both singular defects (domain walls, anisotropic 2D vortices and Bloch points) and continuous ones, including magnetic skyrmions \cite{skyrme-61, skyrme-62, bogdanov-89, muhlbauer-09}. For the first time, skyrmions were studied by T. Skyrma in nuclear physics as topologically nontrivial configurations of the baryon field \cite{skyrme-61, skyrme-62} and similar structures were predicted in magnetic systems by A.N. Bogdanov and D.A. Yablonsky \cite{bogdanov-89}. Later MSs were observed in MnSi \cite{muhlbauer-09} as well as in other materials \cite{romming-15, nayak-17, kanazawa-17, grenz-17}. In the last decade, the experimental progress \cite{yang-15, moreau-16} has made it possible to create and study magnetic skyrmions in nanowires and thin magnetic films. In this work we will study two-dimensional MSs realized in thin films.

\begin{figure}
\begin{center}
\includegraphics[width=0.5\textwidth]{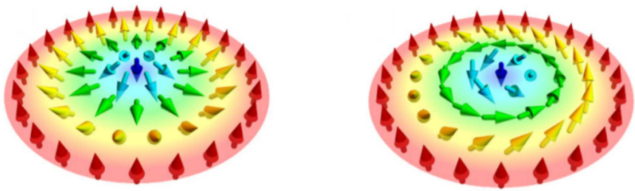}
\caption{Adopted from work \cite{everschor-sitte-18}. Examples of 2D magnetic skyrmions of the Néel (left) and Bloch (right) types. The ordering of the magnetic moments outside the boundaries of the structures coincides with the direction of the moments at the boundary. The direction of magnetization in the center of the structure is opposite to the direction at the boundary.}\label{fig-1}
\end{center}
\end{figure}

Two-dimensional magnetic skyrmions are vortex-like distributions of magnetic moments in the $\mathbb {R}^2$ plane (see, for examples Fig. \ref{fig-1}). In the center of the vortex the direction of the magnetic moment is opposite to the direction at the skyrmion boundaries. For the slowly (compared to the atomic scale) varying magnetic profile we will use the notation $\langle {\bf S}_f \rangle \to {\bf m}({\bf r})$ where the magnetization field ${\bf m} \in \mathbb{S}^2$ is the continuous functions of ${\bf r} \in \mathbb{R}^2$. In practice, the normalization $|{\bf m} ({\bf r})|=1$ is usually used.

Practical interest in MSs is mainly associated with their topological stability: despite the nano- or micrometer vortex scale, MSs are stable to defects and temperature fluctuations. At present, this stability underlies numerous schemes for using MSs in logic \cite{zhang-15, zazvorka-19} and memory \cite{yu-16, miller-17} devices. In particular, methods for recording and reading magnetic information by creating and moving MSs by spin-polarized current \cite{yu-16, miller-17} are proposed. This idea is based on the assumption that MSs do not deform into other magnetic structures upon creating and reading information.

Mathematically, the possibility of continuous deformation of various magnetic structures is related to the homotopy theory \cite{nakahara-03, schwarz-93}. Two magnetic configurations are called topologically (homotopically) equivalent if there is a possibility of their continuous deformation into each other without overcoming the infinite energy barrier. On the contrary, two configurations are topologically nonequivalent if such a continuous deformation is impossible. Then, the magnetic distribution exhibits "topological stability" if it is topologically nontrivial, i.e. if it cannot be deformed into a spatially homogeneous distribution belonging to the trivial homotopy class.

The given homotopy arguments explain the stability of magnetic skyrmions to perturbations. However, the idealized character of these arguments should be noted, as they do not take into account the discreteness of the magnetic structure, boundary effects, and quantum fluctuations. These factors lead to the finite energy barrier necessary to deform the profile $\bf{m}(\bf {r})$ to another homotopy class against the infinite energy barrier in continuous approximation. The qualitative estimates show that this finite barrier can be comparable to the room temperature \cite{uzdin-18}.

In this paper, we consider isotropic magnetic structures with $ |{\bf m} ({\bf r})| = 1 $ on a two-dimensional plane. This means that the configuration space of the magnetic order parameter ${\bf m} ({\bf r})$ can be mapped on the sphere $\mathbb{S}^2$. The two-dimensional plane $\mathbb{R}^2$ is also compactified to $\mathbb{S}^2$. Thus, various magnetic configurations are characterized by the mapping $\mathbb {S}^2 \to \mathbb{S}^2$, and the homotopy group of these configurations is the group $\pi_{2} \left (\mathbb {S}^2 \right) \sim \mathbb{Z} $. The latter is isomorphic to the group of integers $ \mathbb{Z}$.

To establish an exact correspondence between the fields ${\bf m} ({\bf r})$ and elements of the homotopy group $ \pi_{2} \left (\mathbb{S}^2 \right)$, we use the concept of the degree of the mapping $Q$. The latter is also called "winding number", "topological index" or "topological charge". For the considered mapping $\mathbb {S}^2 \to \mathbb{S}^2$ it has the form \cite{nakahara-03, schwarz-93}:
\begin{equation}
\label{Q}
Q = \frac{1}{4\pi}\int_{-\infty}^{\infty}\int_{-\infty}^{\infty}
\left( {\bf m} \cdot \left[\,\frac{\partial {\bf m}}{\partial x}\times \frac{\partial {\bf m}}{\partial y} \,\right] \right)\,dx\, dy.
\end{equation}
This characteristic takes the integer values $ Q \in \mathbb{Z}$ and its geometric meaning is the number of times the vector $ {\bf m}$ winds upon the mapping $ \mathbb{S}^2 \to \mathbb{S}^2$.

\begin{figure}[htbp]
	\includegraphics[width=0.5\textwidth]{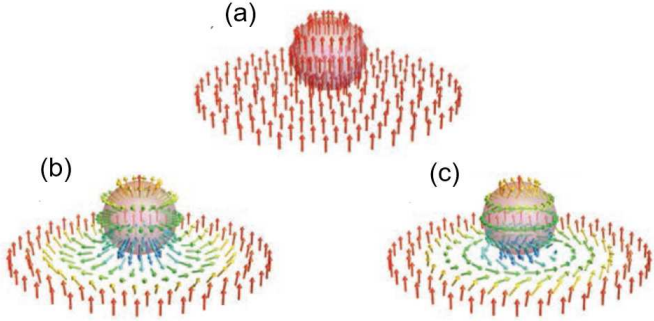}
\caption{Examples of various magnetic structures ${\bf m} ({\bf r})$ and demonstration of their connection with the elements of the homotopy group $\pi_ {2} \left (\mathbb{S}^2\right) $ by means of stereographic projection. (a) Homogeneous magnetic distribution with $Q = 0$. (b) and (c) Skyrmions of the Néel and Bloch type with $Q = -1$.} \label{fig-2}
\end{figure}

The examples of magnetic structures with the topological indices $Q = 0$ and $ Q = -1 $, as well as their stereographic projections, are shown in Fig. \ref{fig-2}. In Fig.\ref{fig-2} (a) the uniform distribution of magnetic moments is demonstrated. On the sphere $\mathbb{S}^2$ this field can be collected into a point. Thus, the ferromagnetic ordering belongs to the trivial class with $Q = 0$. Fig.\ref{fig-2} (b) and (c) shows similar mappings for the skyrmions already discussed in Fig.\ref {fig-1}. These distributions cannot be continuously deformed into a homogeneous structure, which corresponds to the well-known mathematical statement about ``the impossibility of combing a hedgehog without creating a cowlick''. Thus, the distributions $ {\bf m} ({\bf r})$ belong to the nontrivial homotopy class and have the topological charge $Q = -1$.

The energy functionals which support the local or global minima corresponding to MSs should include competing interactions. These interactions should, on the one hand, tend to collinear magnetic ordering and, on the other hand, break the chiral symmetry \cite {gobel-21}. For two-dimensional magnetic structures, the widely used functionals are of the following form:
\begin{equation}
\label{En_func}
E = E_{ex}+E_{an}+E_{Ze}+E_{DM}\, ;
\end{equation}
where $E_{ex}$, $E_{an}$, $E_{Ze}$ is the exchange interaction, energy of the single-ion anisotropy and Zeeman splitting, respectively. These interactions take the form:
\begin{align}
\label{En_coll}
E_{ex}=\mathcal{J}\int_{\mathbb{R}^2} |\nabla m|^2 dS;~~E_{an}=\mathcal{K}\int_{\mathbb{R}^2} (1-m_z^2)dS;\nonumber \\
E_{Ze}=\mathcal{B}\int_{\mathbb{R}^2} (1-m_z)dS,
\end{align}
where $dS = dx\wedge dy$. Their role in the energy functional is to create a tendency towards collinear ordering along the $z$ axis perpendicular to the plane.

The Dzyaloshinsky-Moriya interaction $ E_ {DM} $ is widely used as the chiral interaction in the theory of magnetic skyrmions. If the Néel and Bloch skyrmions are realized as a (meta-) stable configuration, then the functional $ E_{DM} $ has the form: $$ E_{DM} = \mathcal{D} \int_{\mathbb{R}^2} \omega dS, $$ where the function $\omega$ is written as
\begin{enumerate}
    \item for the Néel case
    \begin{eqnarray}
    \label{En_DM}
    \omega = m_{z}\frac{\partial m_x}{\partial x} - m_{x}\frac{\partial m_x}{\partial x} + m_{z}\frac{\partial m_y}{\partial y} - m_{y}\frac{\partial m_z}{\partial y},
    \end{eqnarray}
    \item for the Bloch case
    \begin{eqnarray}
    \label{En_DM2}
    \omega = m_{y}\frac{\partial m_z}{\partial x} - m_{z}\frac{\partial m_y}{\partial x} - m_{x}\frac{\partial m_z}{\partial y} + m_{z}\frac{\partial m_x}{\partial y}.
    \end{eqnarray}
\end{enumerate}
It should be noted that $ E_{DM} $ is not the only one breaking the chiral symmetry functional. In particular, this symmetry can be broken spontaneously \cite{murakami-03, sinova-04}.

\subsection{Analytical ansatz for the magnetic profiles of skyrmions}
\label{sec:3.2}

For the above mentioned, skyrmion-type equilibrium magnetic configurations can be obtained by solving the Euler-Lagrange equations \cite {buttner-18, wang-18}. However, no analytical solution of these nonlinear equations in the 2D case is yet available. Instead, analytical ansatz \cite{buttner-18, wang-18, romming-15} are used for the theoretical description of the MS properties such as topological superconductivity. For the magnetic skyrmions with $Q = \pm 1$ a few ansatz have been proposed. The general form of the ansatz is
\begin{equation}
\label{sk_param}
m_x = \sin \Theta \cos \Phi,~~
m_y = \sin \Theta \sin \Phi,~~
m_z = \cos \Theta.
\end{equation}
Here, $\Phi = n\varphi + \psi$, where $\varphi \in (\, 0, \, 2 \pi \,] $, $n \in \mathbb{Z}$, and the parameters are $\psi = 0$ for the case of the Néel skyrmion, and $\psi = \pi/2 $ for the Bloch skyrmion. The function $\Theta (r) $ describes the radial behavior of the skyrmion profile presented in Fig. \ref{fig-1} and has the form \cite{buttner-18, wang-18, romming-15}
\begin{equation} \label{theta}
    \begin{split}
&\Theta(r,\,R,\,w) = \Theta_{DW}(-r-R,\,w) + \Theta_{DW}(r-R,\,w) = \\
&= 2\arctan{\left( e^{\left({-r-R} \right)/{w}} \right)} + 2\arctan{\left( e^{\left({r-R} \right)/{w}} \right)} = \\
&=\arcsin{\left(\tanh{\left(\frac{r-R}{w}\right)}\right)}+\arcsin{\left(\tanh{\left(\frac{r+R}{w}\right)}\right)} + \\
&+ \pi = 2\arctan{\left(\frac{\cosh\left(R/w \right)}{\sinh\left(r / w \right)} \right)}.
\end{split}
\end{equation}
Here, the parameter $R$ describes the distance from the center of the MS to its boundary. It can be related to the radius of the skyrmion boundary $\mathcal {R}$, for which $m_z(\mathcal {R})=0 $ using formula \cite{wang-18}
\begin{equation}
\label{Sk_Radii}
\mathcal{R}=w\, \arcsin{\left(\frac{\exp\left(2R/w\right)+1}{2\exp\left(2R/w\right)} \right)}.
\end{equation}
Such a definition is convenient in view of its unambiguity. The parametrization (\ref{sk_param}), (\ref{theta}) was used in \cite{wang-18} to build an analytical theory of the skyrmion radius. Also, the profile described by such a parameterization was compared to the profiles obtained in physical and numerical experiments, and it demonstrated excellent agreement. In addition, the privilege of the substitution of $\Theta(r)$ in the form (\ref{theta}) is to use the functions $\Theta_{DW}(r)$ which is the exact solution of the Euler-Lagrange equations for domain walls in 1D chiral magnets. In this case, the parameter $w$ describes the domain wall width. If the skyrmion domain walls do not overlap and the magnetic moments flip $p$ times in radial movement, expression (\ref{theta}) for $\Theta(r)$ is generalized by replacing $$ \Theta(r, R, w) \to \Theta(r, w, R_1, \ldots, R_p) = \sum_{i = 1}^{p}\Theta(r, R_i, w ).$$
In addition to the accurate and quite complicated skyrmion parameterization (\ref{sk_param}), (\ref{theta}) a less accurate and more simple analytical ansatz is also used. The first type of the latter is the so-called linear parameterization
\begin{equation} \label{theta_lin}
\Theta \to \Theta_l = \left\{\begin{array}{*{20}{c}}
\pi\,,~~~~ r<r_0\,; \\
~~\left(r - r_0\right)/R,\,~~~r_0<r<r_0+pR; \\
~~~~~~2\pi\,,~~~~ r>r_0+pR\,.
\end{array}\right.
\end{equation}
Moreover, the so-called exponential parameterization of simple skyrmions, defined by the expression
\begin{equation} \label{theta_exp}
\Theta \to \Theta_e = \pi\exp\left(-\frac{r}{R}\right)
\end{equation}
is used. Note that the parameters $ R $  characterizing the skyrmion radius lead to different skyrmion sizes, with other things being equal.

The above parameterizations correspond to the direction of the field $m_ {z} = \mp1$ in the center (far from the boundary) of the skyrmion, and to the topological index of the structure $Q = -1$. In general, the relationship between the value of the topological index (\ref{Q}) and the parametrization functions (\ref{sk_param}), (\ref{theta}) has the form
\begin{equation}
\label{Q2}
Q = \left[\frac{\Phi(\varphi)}{2\pi}\right]_{\varphi=0}^{\varphi = 2\pi}\left[\frac{-\cos\Theta(r)}{2}\right]_{r=0}^{r \to \infty}.
\end{equation}

The expressions for the energy functionals for the above parameterization take the form
\begin{equation}\label{EF2}
    \begin{split}
&E_{ex}=2\pi \mathcal{A} \int_0^{\infty} \left[\left(\frac{d\Theta}{dr}\right)^2+\frac{\sin^2\Theta}{r^2}\right]r dr,\nonumber\\
&E_{DM}=2\pi \mathcal{D} \int_0^{\infty} \left[\frac{d\Theta}{dr}+\frac{\sin 2\Theta}{2r}\right]r dr, \nonumber\\
&E_{an}=2\pi \mathcal{K} \int_0^{\infty} \sin^2 \Theta r dr, \nonumber\\
&E_{Ze}=2\pi \mathcal{B} \int_0^{\infty} (1-\cos \Theta) r dr.
\end{split}
\end{equation}

It is important to note that the described parameterizations refer only to the cases of the so-called simple skyrmions with $ |Q| = 1 $. Currently, a zoo of skyrmion-type structures (partially described in Sec.\ref{sec:3.5}) is open. In Sec.\ref{sec:3.4} we consider antiferromagnetic skyrmions which can be parameterized by (\ref{sk_param}) and (\ref{theta}) assuming that the neighboring magnetic moments have opposite directions when moving in the radial direction from the skyrmion center.

\subsection{Majorana modes in single skyrmions}
\label{sec:3.2a}

The topological stability, locality, and controllability create attractive prospects for using MSs as carriers of MBSs. Therefore, contrary to 1D superconducting nanowires, the braiding of the Majorana modes localized on MSs does not require the creation of complex $ T- $, $X-$ and $ Y- $ junctions. Recently, 2D superconductor / chiral magnet hybrid structures have been proposed as one of the systems supporting MMs \cite{chen-15, yang-15}. In these systems, the rotational symmetry in the real space is broken due to the $s-d(f)$ - exchange interaction between itinerant electrons and magnetic vortex. Thus, the localization area of MMs is related to one of MS. This makes it potentially possible to realize the MM braiding by moving MSs using the mechanisms described above. The Hamiltonian of the hybrid structure has the form
\begin{eqnarray}\label{HBdG}
H &&= \int_{S} \hat{{\rm \Psi}}^{+}({\bf r})\,\mathcal{H}\,\hat{{\rm \Psi}}({\bf r})\,dS,\\
\mathcal{H} &&= \left(-\frac{{\bm \nabla}^2}{2m} - \mu \right)\tau_z + J\,{\bm \sigma \cdot \bf m(r)} + \left(\Delta({\bf r})\,\tau_{+} + h.c.\right). \nonumber
\end{eqnarray}
Here, $\hat{{\rm \Psi}} = \left(c_{\uparrow}, \, c_{\downarrow}, \, c^{+}_{\downarrow}, \, -c^+_{\uparrow} \right) $ is the field operator in the Nambu representation; $ {\bm \sigma}$ and ${\bm \tau} $ are the vectors of the Pauli matrices acting in the spin and electron-hole spaces, respectively; $\mu$ and $m$  are the chemical potential and the effective mass of the itinerant electrons, respectively; $J$ is the parameter of the $s-d (f)$ coupling between the itinerant electrons and the localized magnetic moments of an axially symmetric skyrmion. The spatial dependence of the skyrmion profile ${\bf m(r)}$ is described by parametrization (\ref{sk_param}). The function $\Delta({\bf r})$ describes the spatial dependence of the superconducting order parameter which is proximity induced by the bulk superconductor. With regard to further consideration of type-II superconductors with vortices, we define this function in the general form\begin{eqnarray}\label{Delta_vortex}
~~~~~~~~~~~~~~~~~
\Delta({\bf r})=\Delta_0\,e^{ib\varphi}\left(1-e^{-r/R_v}\right).
\end{eqnarray}
In the last expression, $\Delta_0$ is the superconducting order parameter, $b$ and $R_v$ characterizes the vorticity and radius of the SC vortex. In this section, it will be assumed that $b=0$, $R_v \to \infty$. The effects of the finite $b$ and $R_v$ will be discussed in Sec.\ref{sec:3.3}.

The energies of the single-particle excitations $\varepsilon$ and single-particle wave functions $\Psi({\bf r})$ are determined from the solution of the eigenproblem for the BdG Hamiltonian
\begin{eqnarray}\label{BdG_skyrm}
~~~~~~~~~~~~~~~~~~~~~~
\mathcal{H}\,\Psi_{m}({\bf r})=\varepsilon_m\,\Psi_{m}({\bf r}).
\end{eqnarray}
In Refs.\cite{yang-16, rex-19} it is shown that the Hamiltonian $\mathcal{H}$ commutes with the modified orbital momentum operator\begin{eqnarray}\label{J_gen}
~~~~~~~~~~~~~~~~~~~
J_z = -i\frac{\partial}{\partial \varphi}+\frac{n}{2}\sigma_z-\frac{b}{2}\tau_z,
\end{eqnarray}
and therefore, the solution of the eigenproblem can be factored in the polar coordinate system in the form
\begin{eqnarray}\label{Gamma_polar}
~~~~~~~~
\Psi_m({\bf r}) \to \tilde{\Psi}^l_m(r,\varphi) =  e^{\varphi\left(l - \frac{n}{2}\sigma_z + \frac{b}{2}\tau_z\right)}\,\Psi^{l}_{m}(r),
\end{eqnarray}
where $\Psi_m^{l}(r)=[u^l_{\uparrow}(r),\,u^l_{\downarrow}(r),\,-v^l_{\downarrow}(r), \, v^l_{\uparrow}(r)]$ is the four-component eigenfunctions of the operator
\begin{eqnarray}\label{HBdG_r}
&&\mathcal{H}^l=-\frac{1}{2m}\left[\partial_r^2+\frac{1}{r}\partial_r + \frac{1}{r^2}\left(l - \frac{n}{2}\sigma_z + \frac{b}{2}\tau_z \right)^2 \right]\tau_z-\nonumber\\
&&-\mu\tau_z+J\sigma_z\cos\Theta(r)+
J\sigma_x\sin\Theta(r)+\Delta(r)\tau_x.
\end{eqnarray}
The coefficients $u^l_{\sigma}(r)$ and $ v^l_{\sigma}(r)$ determine the radial components of the particle-like and hole-like wave functions. The periodicity of the latter as related to the variable $\varphi$ allows us to make a conclusion concerning certain properties of MSs hosting MMs. Thus, if $\Psi_m(r, \, \varphi) = \Psi_m(r, \, \varphi + 2\pi)$, then the eigenvalues of $J_z$ lie in the integers, $ l - n / 2 + b / 2 \in \mathbb{Z} $. On the other hand, the solutions of the BdG equations $\Psi^l_m (r, \varphi)$ corresponding to MM should be invariant under the particle-hole symmetry $C \, \Psi^l_m (r, \varphi) = \Psi^l_m (r , \varphi)$, where $C = \sigma_y \tau_y K $, and $ K $ is the complex conjugation. Since $\{\,J_z \, , \, C\,\} = 0$ and $J_z \, \Psi^l_m(r, \varphi) = l \, \Psi^l_m(r, \varphi)$, then $J_z \, C \, \Psi^l_m (r, \varphi) = - l \, C \, \Psi^l_m (r, \varphi) $. Thus, only the solutions $\Psi^0_m (r, \varphi)$ corresponding to $ l = 0 $ can satisfy the particle-hole symmetry, and therefore, they correspond to MBSs. Taking into account the fact that $ l - n / 2 + b / 2 \in \mathbb {Z} $, it follows that the vorticity parameters of the skyrmion and superconducting vortex must satisfy certain conditions for the MBS realization. Therefore, in the homogeneous phase ($b = 0$) $n$ must be even. However, if the bound state of the skyrmion and superconducting vortex is realized, then the sum $n + b$ must be even.
\begin{figure}[htbp]
	\includegraphics[width=0.5\textwidth]{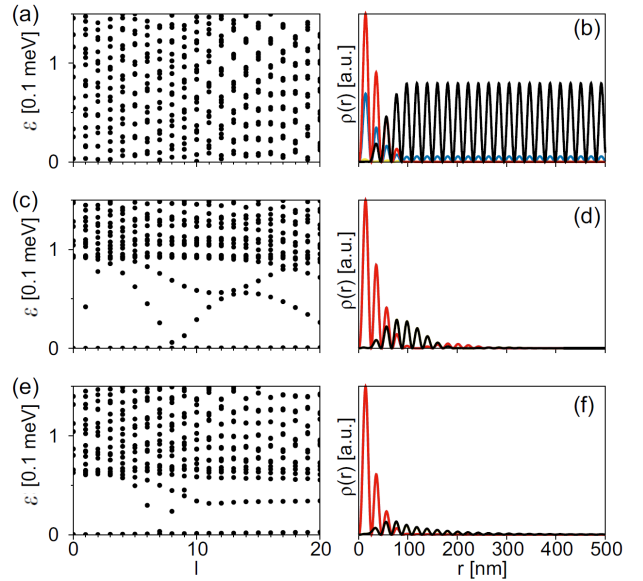}
\caption{Excitation spectrum (left panels) and contributions to the probability density $\rho_m(r)$ of the lowest eigenstate in the $l = 0$ sector (right panels, with the $u^0_{\uparrow}$, $u^0_{\downarrow}$, $v^0_{\uparrow}$, $v^0_{\downarrow}$ components shown in yellow, blue, black, and red, respectively). (a),(b) Skyrmion with $p = 1$. (c),(d) Skyrmion with $p = 10$. (e),(f) Skyrmion with $p = 1$ in the presence of spin-orbit coupling described by $\mathcal{H}_{soc}$ (\ref{H_soc})  \cite{yang-16}.} \label{fig-klinovaja_16a}
\end{figure}

\begin{figure*}[htbp]
\begin{center}
	\includegraphics[width=0.8\textwidth]{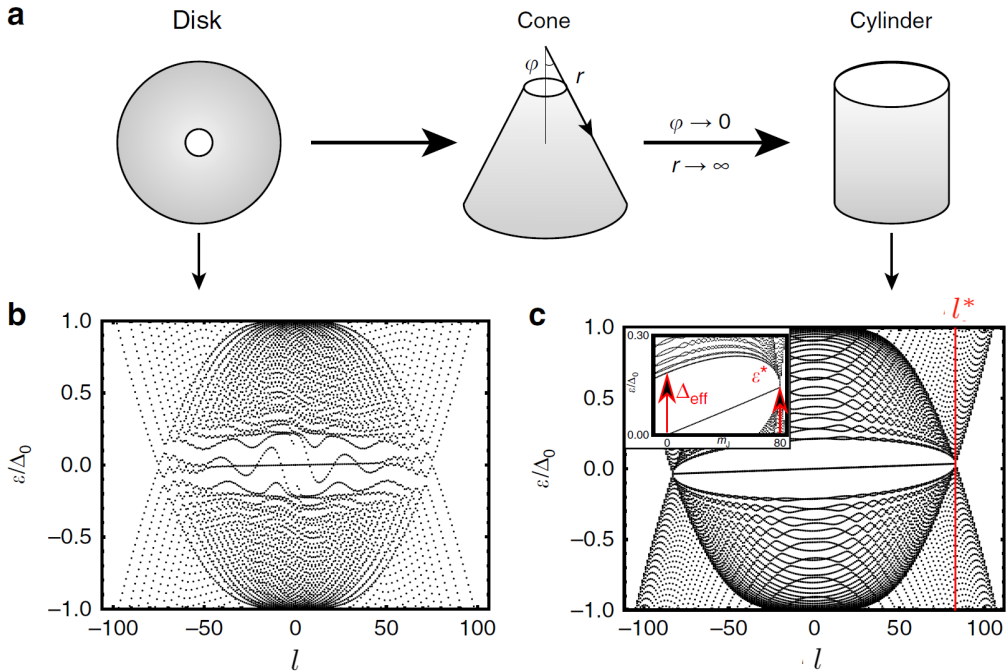}
\caption{ Mapping the skyrmion from a disk to a cylinder (adopted from work \cite{garnier-19}). a) The angle $\varphi$ introduces the mapping: $\varphi = \pi/2$ realizes the disk geometry while the limit $\varphi\to 0$ with $r\to \infty$ and $r \sin \varphi = R_{sk}$ realizes the cylinder geometry, where $r$ is the distance with respect to the tip of the cone. The core is covered by a
white disc for clarity. The excitation spectrum $\varepsilon$ of the original model (b) and the model on the cylinder (c), as a function of the angular momentum quantum number $l$ (the inset shows the same dependencies on a larger scale).} \label{fig-garnier_19a}
\end{center}
\end{figure*}
Thus, in a homogeneous superconductor, MMs can occur only for skyrmions with even $n$. Other conclusions regarding the properties of MMs and MSs can be drawn by considering the properties of the solution $\Psi^{0}_m(r) $ for $l=0$. This problem was solved in Ref.\cite{yang-16} using the linear parameterization (\ref{theta_lin}) when $\Theta(r) \to \Theta(r) + \pi$. Performing a unitary rotation in the spin space described by the operator $U (r) = \exp(i \sigma_y \theta(r)/2)$, the problem is reduced to the 1D one, becoming similar to the problem of the MMs implementation in superconducting nanowires with the spin-orbit interaction. This analogy makes it possible to obtain the conditions of MMs and analytically find their wave functions. Thus, for $n = 2$, MMs can appear only with sufficiently strong exchange interactions between the itinerant electrons and MS $J > \tilde{\mu}^2 + \Delta_0^2$, where $\tilde{\mu}=\mu - \Theta'^2/8m$. In this case, a pair of MMs is realized in the system. One MM (corresponding to the operator $b'$) is localized in the vicinity of the skyrmion center, while the second MM ($b''$) is localized in the vicinity of the skyrmion boundary at $r \approx r_0 + pR$. The characteristic length $\xi$ of MMs $b'$ and $b''$ with the realistic parameters of the system is $\xi \sim R $. This means that weak hybridization of MMs can be achieved only for skyrmions with the large radial vorticity, $p >> 1 $. This additionally restricts the structure of MS hosting MMs.

The excitation spectrum and probability density $$\rho_m (r) = r \, {\overset{+} \Psi} \hphantom {}^{0}_m (r) \cdot \Psi^{0}_m (r),$$ calculated in \cite{yang-16} for different $p$ are shown in Fig. \ref{fig-klinovaja_16a}. Fig.\ref {fig-klinovaja_16a}  (a) and (b) show the dependences for the case $p = 1$. It can be seen that there are no subgap states and MMs hybridize forming quasiparticles with finite energy. In Fig.\ref{fig-klinovaja_16a} (c) and (d) similar dependences are shown for $p=10$. It is seen that a pair of MMs with weakly overlapping wave functions appears in the system. Finally, in the Fig.\ref{fig-klinovaja_16a} (e) and (f) the intrinsic spin-orbit coupling takes into account $\mathcal{H} \to \mathcal{H} + \mathcal{H}_{soc}$. The form of the spin-orbit interaction
\begin{eqnarray}\label{H_soc}
\mathcal{H}_{soc}\sim\{\,\cos\Theta, ({\bf \sigma}\times {\bf p})\cdot {\bf z} \,\} + \{\sin\Theta, ({\bf \sigma}\cdot {\bf p})\}\end{eqnarray}
corresponds to the superposition of the Rashba and Dresselhaus SOCs. Considering Fig.\ref{fig-klinovaja_16a} (e) and (f) one can see that the spin-orbit interaction stabilizes MBSs in the superconducting structures with MS and does not require MS to have a large radius.

In addition to MBSs, subgap states of the chiral, but not of Majorana type are also realized in the superconducting structures with MS. Most of these excitations are localized either near the center, or in the vicinity of MS. Among these, a relatively small number of the states localized near the skyrmion center form a subgap zone with significant dispersion, while a number of the remaining low-energy excitations localized near the edges of MS form a weakly dispersed zone. These excitations have a nonzero orbital momentum and low orbital velocity.

The properties of the subgap states with the weak dispersion were considered in detail in Ref.\cite{garnier-19}. In order to exclude the subgap states localized near the skyrmion center,  conical mapping of the Hamiltonian from the 2D disk to the cylinder surface was performed, see Fig.\ref{fig-garnier_19a}. Consequently, the excitations are localized only near the edges of the system. The mapped Hamiltonian decomposes into the direct sum of the terms $\mathcal{H}^{l}_{eff}$ with the given value of $l$. Inside these subspaces  $\mathcal{H}^{l}_{eff} = \mathcal{H}^{l}_{wire} + \mathcal{H}^{l}_{slope}$. The first term $\mathcal{H}^{l}_{wire}$ describes the one-dimensional TSC with the spin-orbit interaction, which is located in an external magnetic field. This system was well studied, supporting as it does the existence of MMs in the absence of chiral symmetry breaking interactions. The maximum values of the orbital momentum $l$ hosting $\mathcal{H}^{l}_{wire}$ with the gapless spectrum was estimated. Namely, if $|l|<|l^*|$, where $|l^*|=R \sqrt {\mu + \sqrt {J^2- \Delta_0^2}}$, then the term $\mathcal{H}^{l}_{wire}$ has a gapless spectrum.

Thus, the term $\mathcal{H}^{l}_{wire}$ is responsible for the formation of a topologically stable flat band of gapless excitations. The second term $\mathcal{H}^{l}_{slope}$ breaks the chiral symmetry and leads to the transformation of the low-energy states from the Majorana to chiral ones. This is manifested in the appearance of finite-energy excitations with the weak dispersion of the corresponding band. It is important that with the realistic parameters, $\mathcal{H}^{l}_{slope}<<\mathcal{H}^{l}_{wire}$ and therefore, it can be considered as a perturbation. This consideration makes it possible to estimate the linear character of the dispersion of the edge excitation zone as related to $l$. Moreover, the maximum energy of these excitations was estimated as $ \varepsilon^* = (n / R) \sqrt {\mu + \sqrt {J^2 - \Delta_0^2}} $. With the realistic model parameters, this value is small as compared to the effective $p-$wave superconducting gap $\Delta_{eff} $ responsible for the TSCty.

\subsection{Majorana states in a skyrmion-vortex pair}
\label{sec:3.3}

The above arguments show that MMs can be realized in hybrid superconductor / chiral magnet structures with skyrmions of a rather complex morphology. Thus, it is necessary that MS should be characterized by an even azimuthal winding number $n$, and a sufficiently large radial winding number $p$, i.e. it should have a fairly large size. The characteristic energies of the skyrmions of such a morphology were estimated based on the analysis of functionals (\ref{En_func}), (\ref{En_coll}) in Refs.\cite{rex-19, rybakov-19}. In Ref.\cite{rybakov-19}, a discrete version of functional (\ref{Ham_arbQ}) was considered, and the skyrmion state with a large index $p$ was estimated as a metastable one. In Ref.\cite{rex-19} the evolution of magnetic energy, with the skyrmions merging with the formation of structures with even $n$, was considered in the continuum approximation. Both research results showed a significant excitation energy of the skyrmions of complex morphology, corresponding to the large $p$ and even $n$.

\begin{figure}[htbp]
\begin{center}
	\includegraphics[width=0.45\textwidth]{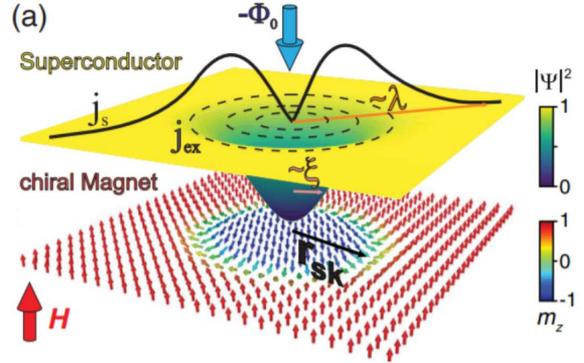}
\caption{ Scheme (adopted from work \cite{petrovic-21}) of a Néel skyrmion with the radius $r_{sk} \equiv R_{sk} \equiv R$ creating an antivortex with the flux $\phi_0 = h/2e$ antiparallel to the external magnetic field. The antivortex currents $j_s$ flow at radii up to $\lambda$. The superconducting order parameter $|\Psi|$ is suppressed over a length $\xi \cong R_v$
in the vortex core.} \label{fig-petrovic_21b}
\end{center}
\end{figure}

One of the ways to overcome this problem was proposed in Ref.\cite{rex-19} while considering the bound state of simple skyrmions and superconducting vortices, see Fig.\ref{fig-petrovic_21b}. Indeed, considering expressions (\ref{Delta_vortex}) and (\ref{J_gen}) it can be seen that the emergence of MBS with $l = 0$ is possible if $n$ and $b$ are odd, in particular $|n|=|b|=1$. It should be noted that the studied superconductor / chiral magnet hybrid structures have attracted great theoretical interest. Thus, magnetic skyrmions can induce Yu-Shiba-Rusinov-type states \cite{pershoguba-16, poyhonen-16} affecting the Josephson current via superconductor / ferromagnet / superconductor junction \cite{yokoyama-15} and they be stabilized by a superconducting dot or antidot located at the top of a ferromagnetic film  \cite{vadimov-18}. The superconducting vortices and skyrmions can also form bound pairs either due to
interplay of proximity effect and spin-orbit coupling \cite{hals-16, baumard-19} or due to their interaction via stray fields \cite{dahir-19, petrovic-21}.

This problem was mainly considered in Ref.\cite{rex-19} using the exponential parametrization of the skyrmion (\ref{theta_exp}), as well as the parametrization (\ref{Delta_vortex}) of the superconducting vortex was performed. In addition, the generalized form of the spin-orbit coupling
\begin{eqnarray}\label{H_soc_gen}
\mathcal{H}_{SOC} = -2i\alpha\tau_ze^{-in'\varphi\sigma_z}\left(-\sigma_y\partial_r + \frac{1}{r}\sigma_x\partial_{\varphi} \right),
\end{eqnarray}
was considered, where $\alpha$ is the spin-orbit coupling parameter. Equation (\ref{H_soc_gen}) describes SOC in which the spin direction winds $n'$ times as it encircles the origin. For $n' = 1$ this type of SOC corresponds to the usual Rashba spin-orbit interaction, while in other cases it varies between the Rashba and Dresselhaus types. In particular, if $n'=2$ $\mathcal{H}_{SOC}$ is identical to SOC considered in \cite{yang-16} in the form (\ref{H_soc}). Moreover, the case with the absence of a skyrmion, but with the presence of a superconducting vortex and Rashba spin-orbit interaction, can be described in terms of $n=0$, $b=n'=1$. In this case, the criterion $n'/2 + b/2 \in \mathbb{Z}$ meets the conditions for the realization of MM in the absence of a skyrmion in the hybrid system. This corresponds to the canonical example for the formation of localized MBS which is a vortex in the 2D spinless p-wave superconductor \cite{alicea-12, beenakker-13, ivanov-01}. If the magnetic skyrmion and generalized spin-orbit interaction, Eq.(\ref{H_soc_gen}), are realized in the system, only the case $n = n'$ corresponds to the conservation of the generalized rotational symmetry and realization of MMs.

\begin{figure*}
  \includegraphics[width=1\textwidth]{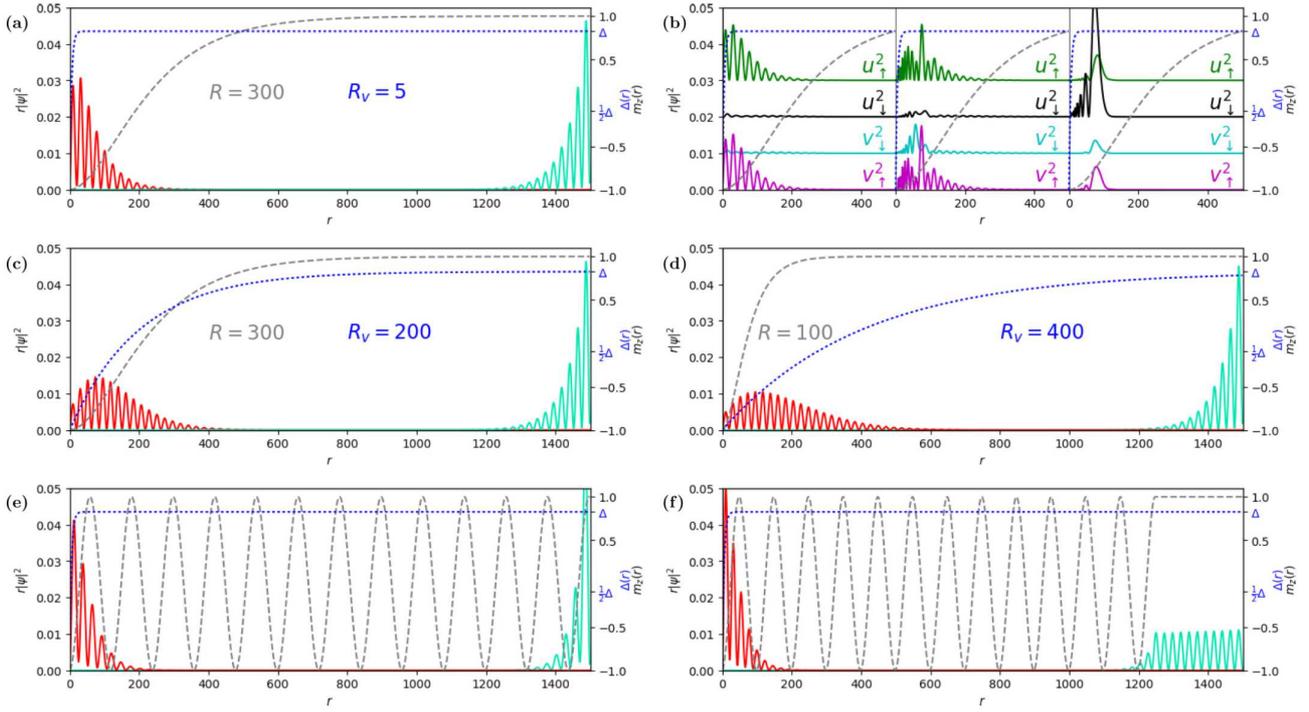}
\caption{Adopted from work \cite{rex-19}. Radial probability density (left y scale) of the inner (red solid line) and outer (turqouise solid line) Majorana modes as well as the radial shape of the skyrmion texture in terms of $m_z$ (dashed grey line, black right y scale) and the profile of the vortex (blue dotted line, blue right y scale). (b) the four components of the inner Majorana bound state from panel (a) (left) in comparison to two examples of localized core states at nonzero energies (i.e., not Majorana bound state) in the same skyrmion-vortex pair (middle: state at $l = -10$, right: state at $l = -12$); (c)-(d) Majorana bound state in the cases shown with an extended vortex. (e)-(f) Majorana bound state in the cases of skyrmions with multiple spin flips. If the outer radius of the skyrmion does not coincide with the system rim the outer mode is delocalized between the outer skyrmion radius and the system rim (case (f)).}
\label{fig_rex-19}
\end{figure*}

The results of the exact diagonalization of Eq. (\ref{HBdG}) [along with Eq.(\ref{H_soc_gen}) whenever $\alpha \neq 0$] are shown in Fig.\ref{fig_rex-19}. The result concerns the case with $n = 1$ and $b = 1$ and the skyrmion is chosen to be of the Néel type. A pair of the Majorana modes is found at $l = 0$, where one of these modes is localized at the core of the skyrmion, whereas the other one is located at the rim of the system (see panels (a) and (b)). Similar to the case of the homogeneous order parameter considered in the previous section, a number of subgap states have been found, some of which are localized at the center of the skyrmion, while other ones at the rim of the system. These additional bound states exist for two reasons. The states localized at the edge are topological ones perturbed by a small interaction breaking the chiral symmetry. The states localized near the skyrmion center are associated with the Yu-Shiba-Rusinov-type ones \cite{pershoguba-16, poyhonen-16}. In Fig.\ref{fig_rex-19} (c) and (d) the impact of the vortex radius is investigated. In both cases, MBSs are still found. However, the localization of the inner MBS becomes weaker with the increasing ratio $R_v/R$. In the panels (e) and (f) the skyrmion-vortex pairs is considered where the effective SOC stems from the winding of the magnetization along the radial direction at $\alpha = 0$. In Fig.\ref{fig_rex-19}(e), the system is restricted to the skyrmion, whereas the skyrmion is embedded in a region of uniform magnetization in the panel (f). The conclusion is that if the skyrmion has a surrounding, the outer mode is not localized at the rim of the system, but it is rather delocalized between the radius of the skyrmion and the rim of the system.

Thus, the bound state of the magnetic skyrmion and superconducting vortex as a source of MBS has several advantages over the homogeneous superconductor. First, this bound state is associated with the well-studied skyrmions with $|Q| = 1$. Secondly, such a system is compatible with the internal spin-orbit Rashba interaction, contrary to a rather exotic type of SOC admissible for the homogeneous superconductor. Thirdly, the pairs of MMs in such a system are localized in the center of the skyrmion and at the rim of the system, which can be of importance in the implementation of MMs braiding. Note that the localization of MMs at the skyrmion radius can be associated with the peculiarities of the linear parametrization. And finally, studies have recently been carried out on the possibility of the experimental realization of the bound state of a magnetic skyrmion and superconducting (anti)vortex.

It is necessary to note some unexplored features of the bound states of the magnetic skyrmion and superconducting vortex. The main features are associated with the mutual influence of MS and the SC vortex on each other. Until recently, there was little research of the issue. However, recently in  Ref.\cite{andriyakhina-21}, a detailed study of the mutual influence of the MS profile and the SC vortex current, as well as energy of the bound state, has been carried out. Three parametrizations of the MS profile were considered: standard parametrization (\ref{theta}) in the form of the superposition of domain walls, exponential (\ref{theta_exp}) and linear (\ref{theta_lin}) parametrization. It is shown that the influence of the skyrmion profile on the SC vortex significantly depends on the type of parametrization, as well as on the skyrmion chirality. In particular, in the case of smooth profiles, the supercurrent decreases monotonically with the increasing distance from the skyrmion center, whereas in the case of the linear parametrization, this dependence is nonmonotonic.
Other features of the interaction between MS and the SC vortex depend on the type of MS (Bloch or Néel one) as well as on the chirality of the Néel skyrmion. Thus, for the Néel skyrmion, depending on the chirality sign, it is energetically favorable for the SC vortex to be either near the skyrmion center or at some distance from it. This dependence is valid both for linear and exponential parametrization. In the case of parametrization (\ref{theta}), the described behavior manifests itself in a certain relationship between the skyrmion radius and domain wall width. For the Bloch skyrmion, it is always energetically advantageous for the center of the SC vortex and MS to be at the same point.

It was also shown in Ref.\cite{andriyakhina-21} that the SC vortex can significantly affect the radius of MS: for Néel MS, the vortex increases the skyrmion radius; for Bloch MS, the SC vortex can both increase and decrease the skyrmion size. However, the intensity of this effect depends on the initial radius of MS. It should be noted that in Ref.\cite{andriyakhina-21} the effect of the vortex on the character of the MS structure is ignored.

Another important problem in the implementation of MBS in superconductor / chiral magnet bilayers is the effect of the ferromagnetic background outside the MS radius on superconductivity. As mentioned above, the amplitude of superconducting pairings can be highly suppressed due to ferromagnetic correlations; thus, the conditions for the implementation of MM can be significantly restricted. One of the possible ways of overcoming the problem is discussed in Sec.\ref{sec:3.4}.

The presented results indicate a large number of unexplored effects which can have significant influence on MM in the bound states of the magnetic skyrmion and superconducting vortex.

\subsection{Topological superconductivity in the skyrmion chain and lattice}
\label{sec:3.4}

\begin{figure}[htbp]
	\includegraphics[width=0.47\textwidth]{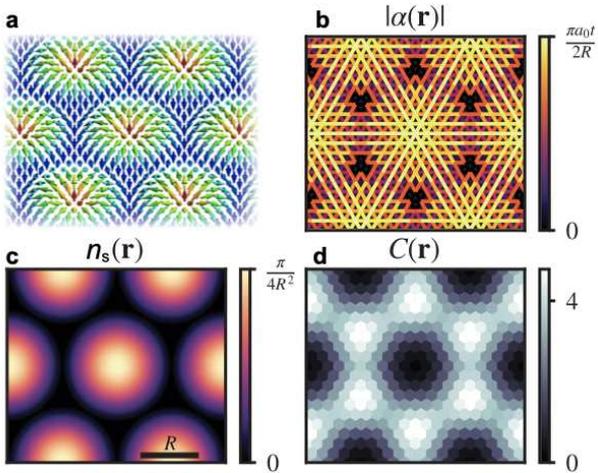}
\caption{System with the magnetic skyrmion lattice. (a) Schematic picture of the skyrmion lattice. Spatial plot of (b) the magnitude of the induced Rashba spin-orbit interaction, $|\alpha({\bf r})|$, (c) the skyrmion number density, $n_s({\bf r})$, and (d) the Chern number density $C({\bf r})$ \cite{mascot-21}.} \label{fig-mascot_21}
\end{figure}

Another way to create superconductor / noncollinear magnet bilayers hosting magnetic excitations with a large topological index $Q$ is to consider skyrmion lattices, ribbons or chains with the proximity-induced superconductivity, Refs.\cite{mascot-21, diaz-21}. In Ref.\cite{mascot-21}, the conditions supporting MMs and their manifestation in the transport properties in a skyrmion lattice were investigated. It is important to note that in this system, superconductivity is not suppressed (or suppressed locally) due to the periodic noncollinear magnetic structure. The main reason for the realization of TSCty in such a system is the effective spin-orbit interaction induced by the noncollinear magnetic structure, whose mechanism was discussed in Sec.\ref{sec:2}. The spatial profile of the skyrmion lattice ${\bf m}({\bf r})$, as well as the effective spin-orbit parameter $|\alpha({\bf r})|$, are shown in Fig.\ref{fig-mascot_21} (a) and (b), respectively. The topological characteristics of the magnetic and electronic subsystems were investigated by analyzing the density of the topological charge $n_s({\bf r})$ and Chern number $C({\bf r})$ defined as
\begin{eqnarray}\label{nr_Cr}
Q = \int_{\mathbb{R}^2} n_{s}({\bf r}) \,dS~,\hphantom{aaa}
C = \int_{\mathbb{R}^2}C({\bf r})\,dS.
\end{eqnarray}
Here, $Q$ is defined by expression (\ref {Q}), and the first Chern number is defined in a standard way
\begin{eqnarray}\label{Chern}
&&C = \frac{1}{2\pi i}\int_{BZ}\text{Tr}\left(\left( P_k \cdot \left[\,\frac{\partial P_k}{\partial x}\,,\, \frac{\partial P_k}{\partial y} \,\right] \right) \right) dk_x \wedge dk_y, \nonumber\\
&&P_k = \sum_{\varepsilon_m({\bf k}) < 0}\Psi_m({\bf r})\,\Psi^+_m({\bf r}).
\end{eqnarray}

The spatial dependences $n_s({\bf r})$ and $C({\bf r})$ are shown in Fig.\ref{fig-mascot_21} (c) and (d). One can see a heterogeneous profile of the parameters characterizing the TSCty. The spatial dependences of ${\bf m}({\bf r})$, $|\alpha({\bf r})|$ and $n_s({\bf r})$ are qualitatively similar, whereas $C({\bf r})$ demonstrates the opposite behavior. The nontrivial spatial behavior of $C({\bf r})$ leads to nontrivial values of the Chern number $C \neq 0$, and the topological phase diagram of the system shows a wide range of the values of $C$ \cite{mascot-21}. Another essential feature of the system is the nonlinear dependence of $|\alpha({\bf r})|$ on the skyrmion radius. Thus, topological transitions can be induced by changing the size of skyrmions by applying an external magnetic field. At the same time, the critical fields required for the topological transitions can be lower than those for many conventional superconductors \cite{mascot-21}. Taking into account experimental observations of skyrmion lattices in chiral magnets, the latter, in the proximity of a conventional superconductor, are considered promising for observing the topological phase transitions. In Ref.\cite{mascot-21} the Josephson scanning tunneling spectroscopy was proposed as an effective tool to visualize the spatial profile of $\Delta({\bf r})$. It is shown that if the superconducting order parameter of the microscope tip is real, then Josephson scanning tunneling spectroscopy allows one to observe the spatial profile of the triplet SC pairing amplitudes inducing TSCty.

When considering the skyrmion lattice in the quasi-one-dimensional geometry, i.e. skyrmion ribbon, in the system, the $ | C | $ pairs of MMs are realized. However, issues concerning the MM manipulation in the skyrmion lattice and ribbons have not been considered and remain open.

\begin{figure}[htbp]
	\includegraphics[width=0.45\textwidth]{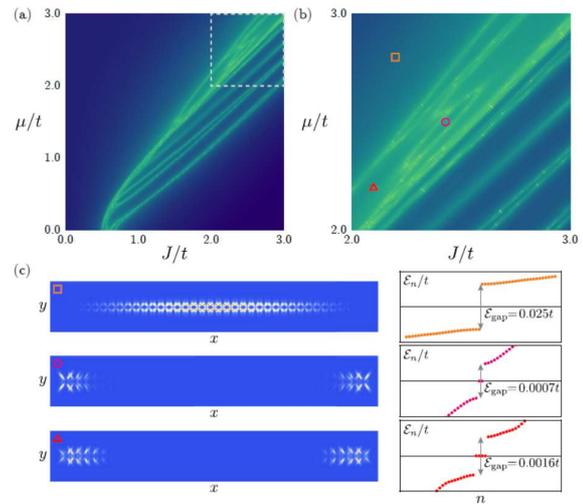}
\caption{Topological phases induced by antiferromagnetic skyrmion chain \cite{diaz-21}. (a) Topological phase diagram as a function of the chemical potential and exchange coefficient. The logarithmically-scaled color code shows the energy gap, $\varepsilon_{0}$, and the green curves denote $\varepsilon_{0}=0$.
(b) Blow-up of the region enclosed by the dashed gray square in (a). The selected phases supporting no Majorana bound states (orange square), two (magenta circle), and four Majorana bound states (red triangle) are highlighted. (c) Probability density of the lowest nonnegative energy state (left) and energy spectrum (right) of the selected phases indicated in (b) for the chain composed
of 37 antiferromagnetic skyrmions.} \label{fig-klinovaja_21}
\end{figure}

MBSs in extremely narrow ribbons, i.e. skyrmion chains, were considered in Ref.\cite{diaz-21}. To avoid the problem of the suppression of superconductivity by ferromagnetic correlations, an antiferromagnetic skyrmion chain (ASC) was considered, see Fig. \ref{fig-klinovaja_21}. The topological transitions and MBSs are also realized in this system.  Moreover, for topological quantum computations, MMs on ASC have advantages over those in 1D non-collinear two-sublattice magnets. The main advantage is a shorter localization length of MBSs on ASC as compared to the case of magnetic chains. The reason for this is the partial delocalization of MMs in the transverse direction of the ribbon along the antiferromagnetic skyrmions. In ASC, MMs are robust to local fluctuations of the chemical potential, $s-d (f)$ exchange coupling and disorder. However, these modes are sensitive to fluctuations of the magnetic profile of skyrmions. This feature can be considered as an additional advantage of MMs on ASC as compared to MMs on magnetic chains, since one can control the spatial position of MMs by distorting MSs. Also, the braiding of MMs can be performed by externally moving MSs by the magnetic tip or spin-polarized current. In Ref.\cite{diaz-21} the device geometry supporting the braiding of MMs was proposed. Also note that another method has recently been proposed for creating skyrmion chains with Majorana modes. A hybrid structure superconductor - ferromagnet hosting a skyrmion lattice was considered, in which the geometry of the superconductor corresponded to the planar Josephson junctions \cite{mohanta-21}. The Majorana modes were also found and studied in such systems.

\subsection{Skyrmions with an arbitrary topological charge}
\label{sec:3.5}

\begin{figure*}
  \includegraphics[width=1\textwidth]{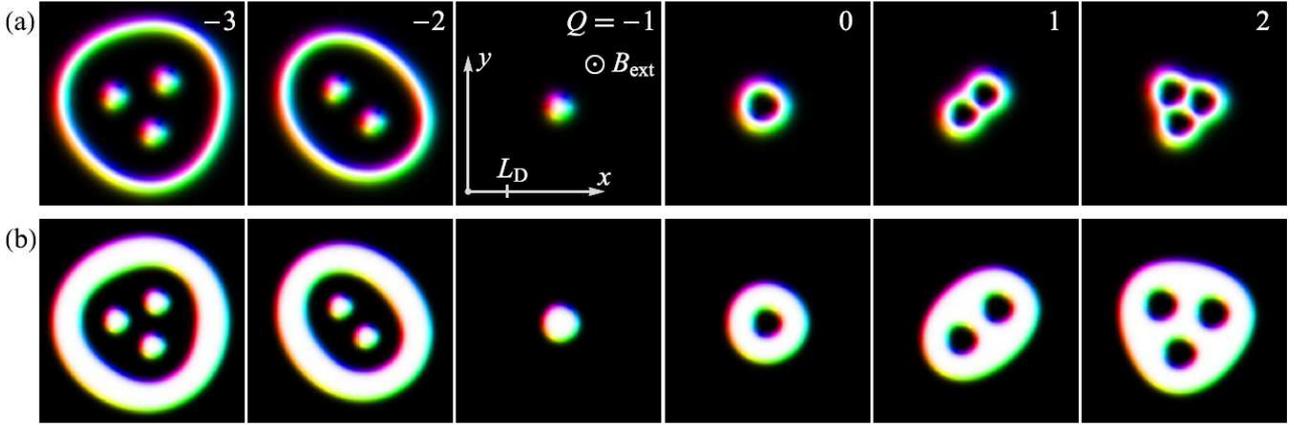}
\caption{Morphology of the stable chiral skyrmions with the topological charges $Q=-3,-2,\ldots, 2$ \cite{rybakov-19}. The top row of the images (a) corresponds to the zero magnetocrystalline anisotropy in the external magnetic field applied perpendicular to the plane. The bottom row of the images (b) corresponds to the case of the uniaxial anisotropy and zero external field. All the images are given in the same scale. The colors show the
direction of the {\bf n} vectors according to the standard scheme: black and white denote the up and down spins, respectively, and red-green-blue show the azimuthal angle with respect to the $x$ axis.}
\label{fig:sk_arbQ}
\end{figure*}

In the conclusion of this section, describing the prospects of the TSCty on MSs, let us note some possible extensions of this research which seem promising to us. In Sec.\ref{sec:3.3}, the possibility of realizing the topological superconductivity in skyrmions with a rather complex morphology is described. Other stable chiral magnetic structures with a large value of the topological index $Q$ are magnetic skyrmions of complex morphology discovered recently in numerical experiments. These studies were carried out by minimizing the discrete version of energy functional (\ref{En_func}) of the form
\begin{equation}
\label{Ham_arbQ}
E=\sum_{\langle i,j \rangle} \big[I{\bf m}_i {\bf m}_j+{\bf D}_{ij}({\bf m}_i\times {\bf m}_j)\big]+\sum_{i}\big[B m_{i,z}+K m_{i,z}^2\big].
\end{equation}
Here, $ i $, $ j $ are the indices belonging to the two-dimensional lattice; $\langle \cdot \,, \, \cdot \rangle$ means the summation over the nearest neighbors. If the vectors ${\bf D}_ {ij}$ (the Dzyaloshinsky-Moriya vectors) are directed perpendicularly (along) the bonds connecting the nearest neighbors, then the local minima of the functional correspond to the Néel (Bloch) skyrmions. The energy parameters of the continuous and discrete models are related to each other. Particularly, for the square lattice, this relationship has the form $$\mathcal{B} = \frac{B}{a^2};~~~\mathcal{K} = \frac{K}{a^2};~~~\mathcal{J}=\frac{I}{2};~~~\mathcal{D}=\frac{D}{a},$$ where $a$ is the  lattice parameter and $D =|{\bf D}_{ij}|$. Note that the continual and lattice functionals, (\ref{En_func}) and (\ref{Ham_arbQ}), are equivalent up to the energy of the ferromagnetic ordering.

Among the numerical approaches to minimizing functional (\ref{Ham_arbQ}), let us single out the geodesic nudged elastic band method \cite{lobanov-17}, string method \cite{weinan-02, suess-19} and minimum mode following method \cite{olsen-04, muller-18} which take into account the curvature of the configuration space, allowing us to search for saddle points on energy surfaces with the dimensions greater than $10^6$ and, thus, to investigate micron-scale structures with the atomic resolution corresponding to MS studies. Also, minima (global and local) of the energy functionals are often searched for using the gradient method.

Recently, using these approaches, the existence of the nontrivial MS with arbitrary values of the topological index $Q$ has been predicted \cite{rybakov-19, foster-19, kuchkin-20, kind-21, kuchkin-21}. The configuration with several skyrmions ($|Q|=1$) located inside one skyrmionium ($Q=0$, see Fig. \ref{fig:sk_arbQ}) is found to be stable (from the classical point of view) and to have a sufficiently low excitation energy \cite{rybakov-19}. These multiple skyrmions with embedded in a magnetic skyrmionium are called skyrmion bags. Their spatial magnetic profile does not have any continuous rotational symmetry, but it is often characterized by the discrete rotation symmetry \cite{kuchkin-21}. However, their large sizes, arbitrary values of $Q$, and nontrivial morphological structure could make skyrmion bags promising objects for the implementation of the TSCty. Moreover, structures of this type were found experimentally in liquid crystals, where the stability of the structures with the large values of $ Q $ \cite{kind-21} was shown.

Note that to date the magnetic skyrmion bags are much less studied than the skyrmions with $|Q|=1$ even in the framework of the classical analysis. For example, there is no analytical theory for skyrmion bags in contrast to the study of MSs in Ref.\cite{wang-18}. Therefore, there is no theory of TSCty in skyrmion bags, as well as no theory taking into account quantum effects since such theories require an analytical parameterization of skyrmion bags.

Note that quantum fluctuations in the features of simple MS are currently being investigated, although much less intensively, as compared to the classical analysis of MS. It has already been shown that spin fluctuations can significantly renormalize the spectral and thermodynamic properties of the system in the presence of the noncollinear geometry and strong single-ion anisotropy \cite{lin-13, roldan-molina-15, takashima-16, aristov-16, diaz-16, psaroudaki-2017, doucot-18, derras-chouk-18, psaroudaki-18, psaroudaki-19, vlasov-20, mook-20}. In particular, the manifestation of quantum effects can lead to the quantum spin reduction, as well as to the renormalizations of the ground state energy \cite{roldan-molina-15, doucot-18}, quantum tunneling to the trivial magnetic profile \cite{derras-chouk-18, vlasov-20} and influence on the kinetic characteristics \cite{lin-13, psaroudaki-18, psaroudaki-19, mook-20}. All these effects can affect the main characteristics of the skyrmion: its size, energy, lifetime, and magnon spectrum. However, the discussion of these effects is beyond this review.

In conclusion of this section, we note that the presented theoretical arguments demonstrate that magnetic skyrmions are promising for the realization and braiding of the Majorana modes. The main advantage of MS is their locality and controllability. The latter, in turn, stems from the topological stability of the skyrmion-type magnetic structure. The main problem of the experimental implementation of such systems is the need to create MS with nontrivial properties associated either with the complex morphology of MS, or with the need to create bound states of the magnetic skyrmion, i.e. a superconducting vortex. The next section describes 2D systems of another type, in which the braiding of MMs is also possible, namely higher-order topological superconductors.

\section{Higher-order topological superconductors in 2D}
\label{sec:4}
In recent years, the topological classification of condensed matter has been
supplemented by so-called higher-order topological
insulators (HOTI) \cite{Benalcazar-17}. While the conventional topological insulators
have a gapped bulk spectrum and a gapless edge spectrum, HOTIs
have both bulk and edge gaps, but support gapless states at the
surfaces of codimension two and higher. For example, the second-order
2D HOTIs contain gapless corner states and 3D HOTIs contain gapless
hinge states. As for the case of the conventional (first-order)
topological systems the concept of higher-order topology was
shortly extended to the case of higher-order topological
superconductors (HOTSCs).

In addition to general interest in HOTSC as a new topological class
of matter, particular attention is being paid to these superconductors due to the
possibility of constructing well-localized zero-dimensional
gapless modes in 2D: Majorana corner modes (MCMs). These modes are
necessary for realizing the braiding of the Majorana fermions, which
provides quantum computation. There are several reasons for the
attractiveness of 2D HOTSC for this purpose. Firstly, the 2D
system is necessary for the braiding \cite{nayak-08}. While one has to make 2D
constructions (such as T- \cite{Cheng-16}, X- \cite{Zhou-20} or Y-junctions \cite{Harper-19}) of 1D TSCs to meet this requirement, 2D HOTSCs satisfy this condition by
definition. Secondly, it is rather difficult to construct a pure
one-dimensional system. Meanwhile, any broadening of a 1D chain leads
to the delocalization of the Majorana state in the direction
perpendicular to the chain and changing of the excitation
character from the Majorana-like to chiral one \cite{Potter-10,Sedlmayr-16}. At the same time,
though the zero-energy excitations remain gapped from the bulk
excitation, there appear gapless excitations in the transverse
direction, which are not gapped from the Majorana states. HOTSCs
do not have this disadvantages as MCMs are well
localized and gapped both from the bulk and first-order edge
excitations.

Whereas there are many specific models providing MCMs in 2D
systems, their underlying construction principle is the same.
Generally, MMs (of any codimension) appear at the topological
defects of the system. In 2D HOTSC this defect is a domain
wall on the edge of the system. To obtain such a wall one needs to
take a system containing gapless edge excitations protected
by some symmetries and add a perturbation, which breaks one of the
symmetries and opens the gap in the spectrum of the conventional edge
excitations. At the same time, the effective mass (Dirac mass) of the
conventional edge excitations must be different for two adjacent
edges. In this case, the corner between two edges will play the role
of the domain wall providing zero-energy excitations located at this
corner.

Although the procedure described above is necessary for the
construction of 2D HOTSC, it is not sufficient. The obtained
zero-energy excitation has to be protected by a symmetry,
otherwise it can be removed from the system by perturbations.
While the topological protection of the gapless excitations in the
conventional TSC is provided by non-spatial symmetries
(time-reversal, particle-hole and chiral symmetries) the
topological protection of HOTSC is provided by crystalline
symmetries. The two most widespread symmetries providing the
topological protection of HOTSC are the inversion symmetry and the
mirror symmetry. The less common are the $C_4$ and $C_2$ symmetries.
Moreover, this symmetry can not only be a single pure crystalline
symmetry, but a combination of symmetries, thus sometimes
their existence is not obvious.

Another feature of HOTSC consists in the fact that while
the topological phase of the conventional TSC is defined only by the
bulk properties, the HOTSC phases can be both
bulk-determined and boundary-determined. For example, one can
first introduce coupling to the conventional TI or TSC to open the
gap on the whole boundary of the system, making it trivial. And
then, perturbations can be added to close this gap at the specific
points on the boundary and to reopen it in such a way that domain
walls appear in pairs from these (high-symmetry) points. In the
latter case, the existence of MCMs depends on the specific boundary
geometry of the system in contrast to the conventional TSC.

In the following sections we will briefly describe the models
proposed for the realization of HOTSC in 2D, discuss their
application for the realization of the braiding procedure and mention the
HOTSCs problems.

\subsection{Topological insulators with superconducting coupling}
\label{sec:32}

\paragraph{Quantum Spin Hall Insulator with superconducting coupling and external magnetic field.}

The general idea of the MCMs construction was described in detail in
\cite{Zhang-20-prr} at an example of a quantum spin-Hall insulator
with the s-wave superconducting coupling and external in-plane
magnetic field. The Hamiltonian of the minimal model in this case
has the form:
\begin{eqnarray}
\label{Ham_QSHI}
&&\mathcal{H}=\mathcal{H}_{QSHI}+\mathcal{H}_{ex}\\
&&\mathcal{H}_{QSHI}=m(\mathbf{k})\tau_z\sigma_z+v\left(\sin k_x s_z\sigma_x+\sin k_y\tau_z\sigma_y\right)-\mu\tau_z,\nonumber\\
&&\mathcal{H}_{ex}=\Delta\tau_y\sigma_y+g\mu_BB\left(\cos\theta\tau_z s_x+\sin\theta s_y\sigma_z\right),\nonumber\\
&&m(\mathbf{k})=m_0-2m\left(2-\cos k_x-\cos k_y\right).\nonumber
\end{eqnarray}
Here, $\mathbf{s}$, $\mathbf{\sigma}$,
$\mathbf{\tau}$ are the Pauli matrices acting on the spin,
orbital and particle-hole spaces with the basis
\begin{equation}
(c_{k\uparrow},c_{k\downarrow},c^{\dag}_{-k\uparrow},c^{\dag}_{-k\downarrow}).
\nonumber
\end{equation}
The Kronecker product symbol between matrices and unitary matrices $s_0$, $\tau_0$ and $\sigma_0$ are omitted, as it was done above. $\pm m(k)$ is the dispersion of two bands with $2m_0$ intersection and $v$ is spin-dependent hybridization induced with spin-orbit coupling. The magnetic field has the value of $B$ and
the angle between its direction and the $x$ axis is $\theta$. The
effective $g$-factor for different orbitals is the same in the
$x$-direction and opposite in the $y$-direction. The proposed
model has only one crystalline symmetry: inversion symmetry.

The edge states of QSHI (\ref{Ham_QSHI}) in the low-energy
limit can be solved analytically. The edge energy bands are
defined as
\begin{eqnarray}
\label{E_QSHI}
&&E_{e\uparrow/\downarrow}(p)=\mp vp-\mu,\\
&&E_{h\uparrow/\downarrow}(p)=\mp vp+\mu,\nonumber
\end{eqnarray}
where $p$ is the momentum along the edge (in the counterclockwise
direction). The corresponding wave functions are
\begin{eqnarray}
\label{Psi_QSHI}
&&\Psi_{e\uparrow,p}=Ae^{ipx'}(e^{\lambda_1y'}-e^{\lambda_2y'})\left(1,-ie^{i\phi},0,0,0,0,0,0\right)^T,\nonumber\\
&&\Psi_{e\downarrow,p}=is_y\Psi^*_{\uparrow,-p},~\Psi_{h\uparrow/\downarrow,p}=\tau_x\Psi^*_{\uparrow/\downarrow,-p},\nonumber\\
&&\lambda_{1,2}=v/2m\pm\sqrt{(v/2m)^2-m_0/m+p^2},
\end{eqnarray}
where the prefactor $A$ is added for the normalization of the wave
function and $\phi$ is the angle between the normal to the edge and
the $x$ direction, $x'$ is the axis along the edge and $y'$ is perpendicular to it. While the edge spectrum of the system is angular
independent, the spinor part of the wave functions depends on
$\phi$. Consequently, the projection of the full Hamiltonian
(\ref{Ham_QSHI}) on the basis (\ref{Psi_QSHI}) will lead to
non-diagonal components, depending both on the magnetic field and
edge orientation. The resulting edge spectrum depends on the angle
between the magnetic field and the normal to the edge:
\begin{eqnarray}
\label{E_QSHI_edge}
&&E(p=0)=\pm\sqrt{\widetilde{B}^2+\widetilde{\Delta}^2\pm2|\widetilde{B}\widetilde{\Delta}|},\\
&&\widetilde{B}=B\sin(\phi-\theta),~~\widetilde{\Delta}=\sqrt{\Delta^2+\mu^2}.\nonumber
\end{eqnarray}

While the magnetic field is small
$\widetilde{B}<\widetilde{\Delta}$, the effective mass of the edge
states are mainly defined by the s-wave superconducting coupling, and
its sign is the same for any direction of the edge. For
$\widetilde{B}>\widetilde{\Delta}$ according to the inversion symmetry, there appear two intervals, in which the effective mass
changes its sign due to the magnetic field. Consequently, 4
domain wall points appear on the boundary of the system at which the sign
of the effective mass changes, whose position is determined as:
\begin{eqnarray}
\label{phi_QSHI}
&&\phi_{1/4}=\theta\pm\arcsin(\widetilde{\Delta}/B),\\
&&\phi_{2/3}=\theta\mp\arcsin(\widetilde{\Delta}/B)+\pi.\nonumber
\end{eqnarray}
These domain walls give rise to MMs in the
fully opened geometry.

\begin{figure}
  \includegraphics[width=0.45\textwidth]{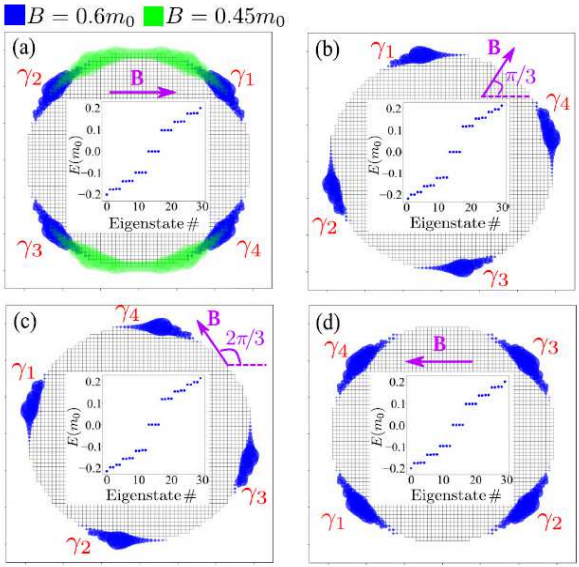}
\caption{Majorana modes with the open disk geometry in QSHI with the s-wave superconducting coupling and external magnetic field \cite{Zhang-20-prr}. The position of the domain walls providing MMs depends on the chemical potential value and magnetic field strength and direction.}
\label{fig:PRR2_OD}
\end{figure}

In the open disk geometry, the angles (\ref{phi_QSHI}) directly
point to the positions of MMs. MMs
appear in two pairs, separated by a angle, which depends only on
the superconducting coupling $\Delta$, chemical potential $\mu$
and magnetic field $B$. Two pairs correspond to each other due to
the inversion symmetry and their position on the disk edge can be
rotated by rotating the in-plane magnetic field (Fig. \ref{fig:PRR2_OD}). The
latter makes the investigated system useful for the braiding and
consequently, for quantum computations.

The obtained MMs remain in the case of a more complicated boundary.
In this case, MMs appears at the corners between two edges,
whose orientation corresponds to different signs of the effective
mass term of the edge excitations. It can be both an advantage
allowing one to reduce the number of MMs from four to two, and a
disadvantage leading to the undesirable creation of additional
MMs at the defects located on the edges or inside the system.

The chosen magnetic field dependence of (\ref{Ham_QSHI}) is not
the only one allowed. The g-factor can be taken to be the same for
different directions. This does not affect the existence of
MCMs in the system \cite{Zhang-20-prb,Wu-20}. But in
this case, MMs cannot be controlled by the direction of
the magnetic field.

The origin of the magnetic field and its structure can also be
more complicated. As it was proposed in \cite{Zhang-19},
QSHI with the conventional s-wave superconducting coupling can be
tuned into the HOTSCty regime by covering it with a bi-collinear
antiferromagnet. According to the feature of the bi-collinear
magnetic structure, the magnetic order along two adjacent edges is
different. It is ferromagnetic in one direction and
antiferromagnetic in the other. The antiferromagnetic order at
the edge does not open the gap in the corresponding edge spectrum,
while the superconducting coupling does. But for the edge with
ferromagnetic ordering along it, there is a competition between the
magnetic and superconducting gaps. In the case when the ferromagnetic
gap dominates, the sign of the effective mass at two adjacent
edges becomes different, providing MCMs in the system.

\paragraph{Quantum Spin Hall Insulator with nodal superconducting singlet coupling.}

While in the previous section, the s-wave superconducting
coupling is used to open the edge excitation gap of the same sign for any
direction of the edge, and then, the magnetic field is used to close and
reopen the gap of the opposite sign in some directions, one can use
the nodal superconducting coupling to open the gap with the opposite
sign on the adjacent edges at once.

The BdG Hamiltonian of the system can be written in the following form \cite{Liu-18,Yan-18}
\begin{eqnarray}
\label{Ham_QSHI_sd}
&&\mathcal{H}=\left(m_0+m_x\cos k_x+m_y\cos k_y\right)\sigma_z\tau_z+\\
&&+\lambda_{so}\left(\sin k_xs_x+\sin k_ys_y\right)\sigma_x\tau_z-\mu\tau_z+\Delta(\mathbf{k})\tau_x,\nonumber\\
&&\Delta(\mathbf{k})=\Delta_0+\Delta_x\cos k_x+\Delta_y\cos k_y,\nonumber
\end{eqnarray}
where $\mathbf{s}$, $\mathbf{\sigma}$ and
$\mathbf{\tau}$ are the Pauli matrices in the spin, orbital
and Nambu spaces with the basis $(c_{k\uparrow},c_{k\downarrow},c^{\dag}_{-k\downarrow},-c^{\dag}_{-k\uparrow})$ and the superconducting coupling is of the extended s-
or d-wave form. The superconducting coupling symmetry is considered to be
the most promising for the practical application: TI grown on
a cuprate-based or iron-based high-$T_c$ superconductor.

Considering the $d_{x^2-y^2}$ case and following the procedure, which was described in detail above, one can easily obtain the effective masses of the edge excitation spectrum:
\begin{eqnarray}
\label{m_QSHI_sd}
&&m_I=\Delta(m_0+m_x+m_y\pm\mu)/m_y,\\
&&m_{II}=-\Delta(m_0+m_x+m_y\pm\mu)/m_x,\nonumber
\end{eqnarray}
for the edges along the $x$ and $y$ directions, correspondingly. If
TI is in the nontrivial topological phase, then the effective
mass at the adjacent sides along $x$ and $y$ has the opposite sign for
$m_xm_y>0$. This results in MMs appearing at the corners of
the system.

\begin{figure}
  \includegraphics[width=0.45\textwidth]{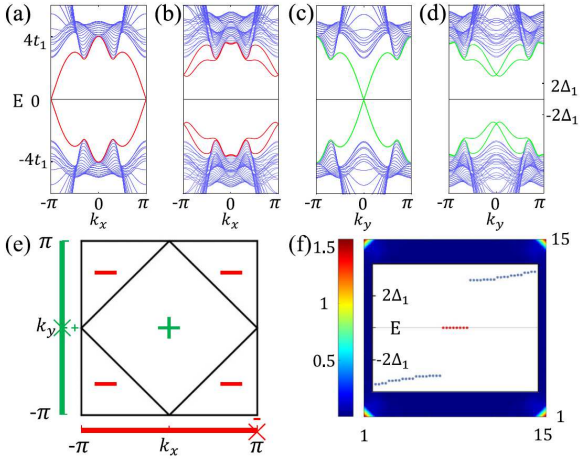}
\caption{The case of the topological insulator with the Rashba spin-orbit interaction and nodal superconductivity \cite{Wang-18-prl}. (a-d) Excitation spectrum of (\ref{Ham_QSHI_sd}) with the edge along the $x$ (a-b) and $y$ (c-d) direction without (a,c) and with (b,d) $s_{\pm}$-wave superconducting coupling. (e) $s_{\pm}$ pairing has different signs for $k_{Dx}$ and $k_{Dy}$, resulting in the appearance of the Majorana corner modes in the square-shaped geometry of the system. (f) Density plot of probability distribution of MCMs in the square-shaped geometry and the lowest excitation spectrum in the inset.}
\label{fig:PRL121}
\end{figure}

If the extended s-wave pairing ($s^{\pm}$-pairing) is considered
in the system, the $m_xm_y<0$ condition for the presence of
MZMs arises, forcing one to have a hopping amplitude with different
signs in the $x$ and $y$ direction in the system.

\paragraph{TI with Rashba spin-orbit interaction and nodal superconductivity.}

While the described above systems have the intrinsic spin-orbit
interaction of the form proposed by Bernevig, Hughes, Zhang \cite{Bernevig-06}
(BHZ), the explicit form of the interorbital spin-orbital
hybridization is not crucial for the construction of MZMs. The
alternative case is the extrinsic interorbital Rashba-like
spin-orbit coupling. Particularly, one can take the minimal model of
the two-orbital model with the interorbital Rashba-like SOC and extended s-wave coupling \cite{Wang-18-prl}:
\begin{eqnarray}
\label{Ham_Rash_sd}
&&\mathcal{H}=\left(h_{TI}(\mathbf{k})-\mu\right)\tau_z+\Delta(\mathbf{k})\tau_x,\\
&&h_{TI}(\mathbf{k})=\left(2t(\cos k_x-\cos k_y)+4t_1\cos k_x\cos k_y\right)\sigma_z+\nonumber\\
&&+2\lambda\left(\sin k_x s_y-\sin k_ys_x\right)\sigma_x,\nonumber\\
&&\Delta(\mathbf{k})=\Delta_0+2\Delta_1\left(\cos k_x+\cos k_y\right).
\end{eqnarray}
Here, $t$ is the nearest-neighbor intraorbital hopping term, $t_1$ is
the next-nearest-neighbor intraorbital hopping, $\lambda$ corresponds
to the interorbital Rashba SOC, $\Delta(k)$ is the
$s^{\pm}$ intraorbital superconducting coupling. The $t_1$ term is
necessary to obtain TI in the absence of superconducting
coupling, otherwise for zero $t_1$ the bulk spectrum has no gap at
the nodal points of the SOC. Notably, for
this, the next-nearest-neighbor hopping can be substituted with
the difference between the on-site energies for two orbitals.

The hopping amplitude $t$, as can be seen from
(\ref{Ham_Rash_sd}), has opposite signs in the $x$ and $y$ directions,
as it was in the previous case. The authors of \cite{Wang-18-prl}
clearly demonstrated this necessity. The sign of the
effective mass is dictated by the sign of the superconducting
coupling at the Dirac points of the TI edge excitation spectrum (Fig. \ref{fig:PRL121}).
Consequently, the rough conditions for the appearance of MCMs result in the sign of
$\Delta(k_{Dx},k_y)$ being opposite to the sign of
$\Delta(k_x,k_{Dy})$ for most $k_x$, $k_y$, where $k_{Dx}$,
$k_{Dy}$ are the momenta of the Dirac point in the edge excitation
spectrum along the $x$ and $y$ direction, correspondingly, in the
cylinder geometry. In the case of the $s^{\pm}$ pairing, this condition
is fulfilled, if one of ${k_D}$ is equal to zero and the
other is equal to $\pi$. It is the case of hopping of the opposite
signs in the $x$, $y$ direction.

As in the case of QSHI with the nodal superconducting coupling,
the $s^{\pm}$ pairing can be replaced with the $d_{x^2-y^2}$ pairing.
In this case, the condition discussed above is fulfilled, when both
$k_{Dx}$ and $k_{Dy}$ are equal to zero or to $\pi$, corresponding to the
hopping amplitudes of the same sign in both directions. The mixed
$s+id$ coupling is also appropriate \cite{Zhu-19}.

\paragraph{Quantum Spin Hall Insulator with odd-parity superconducting coupling.}

The superconducting coupling induced in TI does not have to be of the
singlet form, but can also be of the odd-parity triplet form \cite{Huang-21,Hsu-20}. While the type of the
superconducting coupling and the explicit form of the excitations
change, the situation, in general, remains the same: the
superconducting coupling leads to the opposite-sign masses of the edge excitations at the adjacent edges,
providing MMs on the domain walls, located at the corners.

\subsection{Topological superconductor with perturbations}
\label{sec:33}

\paragraph{$p\pm ip$ TSC with the in-plane magnetic field.}

Previously, we discussed the possibility to construct MMs in TI by opening the conventional edge excitation gap
due to the superconducting coupling and magnetic field. But one can
also start with the helical TSC and add a perturbation, which will open
the edge spectrum gap with the opposite sign for the adjacent edges. One
of the implementations of this approach is the combination of
$p+ip$ and $p-ip$ TSC. In particular, one can take $p\pm ip$
TSC with the $p+ip$ pairing for the spin-up electrons and $p-ip$ pairing
for the spin-down electrons and then, add the in-plane Zeeman field \cite{Zhu-18}:
\begin{eqnarray}
\label{Ham_pip_h}
&&\mathcal{H}=\varepsilon(\mathbf{k})\tau_z-\Delta\left(\sin k_x s_x-\sin k_y s_y\right)+\mathbf{B}\mathbf{\sigma},\\
&&\varepsilon(\mathbf{k})=\mu-2t(\cos k_x+\cos k_y).\nonumber
\end{eqnarray}

In the absence of the magnetic field, Hamiltonian
(\ref{Ham_pip_h}) provides the gapless first-order surface
excitations, propagating in the opposite directions along the edge for
the opposite spin projections. The Zeeman field breaks the
time-reversal and $C_4$ symmetries of the system, opening the gap
in the edge excitation spectrum. The corresponding mass term
depends on the direction of the in-plane magnetic field and on the
orientation of the edge: $m_{eff}=B\cos(\theta+\phi)$, where
$\theta$ and $\phi$ are the angles of the magnetic field
direction and edge normal direction, correspondingly.

The mass term vanishes at two points connected with the inverse
symmetry around the whole surface in the open disk geometry,
leading to the presence of two MMs, whose positions can be
controlled by the direction of the magnetic field. In the square-shaped
geometry, MMs appear at the corners between the edges with
the opposite mass sign in accordance with the general rule of MMs at
the domain walls.

\paragraph{Helical TSC with magnetic field}

While the model of $p\pm ip$ TSC is the simplest, it is not
the only one. Another suitable suggestion is the $\pi$-junction of two 2D
layers with the opposite SOC term \cite{Volpez-19,Plekhanov-21}:
\begin{eqnarray}
\label{Ham_helical_h}
&&\mathcal{H}=2t\left(2-\cos k_x-\cos k_y-\mu\right)\tau_z+\Gamma\tau_z\sigma_x+\\
&&+\lambda\left(\sin k_ys_x-\sin k_x\tau_zs_y\right)\sigma_z+\Delta\tau_ys_y\sigma_z.\nonumber
\end{eqnarray}
Here, again $\mathbf{s}$, $\mathbf{\tau}$ are the Pauli matrices in the spin and particle-hole spaces with the basis $(c_{k\uparrow},c_{k\downarrow},c^{\dag}_{-k\uparrow},c^{\dag}_{-k\downarrow})$, $\mathbf{\sigma}$ are Pauli matrices acting on the subspace of the top/bottom layer, $\Gamma$ is the inter-layer spin-conserving hopping amplitude.

For $\Gamma<\Delta$ the system is trivial and for $\Gamma>\Delta$
it is in the topological phase with the helical first-order edge
excitations. The application of the external magnetic field $B$ to
this system in the topological phase opens the gap in the edge
excitation spectrum $m_{eff}=B\cos(\theta-\phi)$. In the manner
identical to the previously described case of $p\pm ip$ TSC, this leads
to the presence of two MZMs in the fully opened geometry.

\subsection{Dirac semimetal/nodal metal with $p+ip$ superconducting coupling}
\label{sec:34}

Here, it is not necessary to start with an insulating normal state, as in the previous cases.
Another possibility to construct MMs \cite{Tiwari-20,Ahn-20,Roy-20,Wang-18-prb}  is to take a double mirror
Dirac semimetal with a weak $p+ip$ superconducting interorbital
coupling:
\begin{eqnarray}
\label{Ham_DSM}
&&\mathcal{H}=t\left(\cos k_x\sigma_x\tau_z+t\cos k_y\sigma_y\right)+\\
&&+\Delta\left(\sin k_x\sigma_x\tau_x+\sin k_y\sigma_x\tau_x\right).\nonumber
\end{eqnarray}
At $t=\Delta$ limit, it can be rewritten in terms of the Majorana operators:
\begin{eqnarray}
\label{Ham_DSM_maj}
&&\mathcal{H}=-2it\sum_{mn}\left(\gamma^2_{m,n}\gamma^1_{m+1,n}+\gamma^4_{m,n}\gamma^4_{m+1,n}-\right.\\
&&\left.-\gamma^2_{m,n}\gamma^4_{m,n+1}+\gamma^1_{m,n}\gamma^3_{m,n+1}\right)\nonumber,
\end{eqnarray}
where Majorana operators and electron annihilation operators are connected through the relations:
\begin{eqnarray}
\label{Maj_DSM_form}
&&c_{\sigma,m,n}=\left(\gamma^1_{m,n}+i\gamma^2_{m,n}\right)/\sqrt{2},\nonumber\\
&&c_{\overline{\sigma},m,n}=\left(\gamma^3_{m,n}+i\gamma^4_{m,n}\right)/\sqrt{2}\nonumber
\end{eqnarray}
with $\sigma=\uparrow$ for odd $m$ and $\sigma=\downarrow$ for even $m$.

The described system is a 2D analog of the Kitaev
chain. While the Majorana operators in the bulk and at the edge are connected in fours and pairs, correspondingly, and yield the gapped bulk and edge excitation
spectrum, the uncoupled corner-localized Majorana operators
provide zero-energy corner modes. As in the case of the Kitaev chain,
MZMs remain in the system away from the special parameter
point until the gap in the spectrum vanishes.

\subsection{Nanowire-based HOTSC}
\label{sec:35}

While the previous method started with the 2D topological systems, the
second-order TSC can be obtained by coupling of 1D topological
superconducting Rashba nanowires. These nanowires are strongly
coupled in pairs, and the obtained two-wire systems are weakly
coupled, to form a 2D system. The latter system can be implemented
in the direction parallel to the pairwise coupling or in the
direction perpendicular to it.

\paragraph{Two-layered Rashba nanowire system.}

\begin{figure}
  \includegraphics[width=0.45\textwidth]{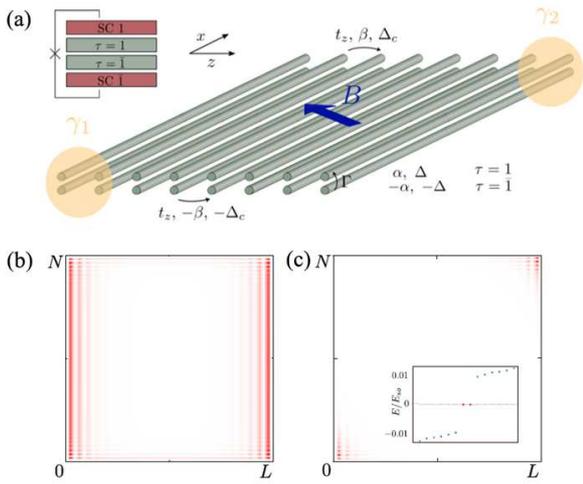}
\caption{(a) Two-layered setup constructed with the Rashba nanowires \cite{Laubscher-19}. (b) In the absence of the magnetic field, the system is the helical topological superconductor with the gapless Majorana modes propagating along the edges. (c) In the presence of a small magnetic field it transforms to higher-order topological superconductor with Majorana corner modes located in two opposite corners of the device.}
\label{fig:PRR1_NW1}
\end{figure}

One can make two layers of the parallel Rashba nanowires with
The opposite Rashba field direction and $\pi$ superconducting coupling
phase difference for different layers (Fig. \ref{fig:PRR1_NW1}). The inter-layer coupling
between the nearest wires is supposed to be strong, while the intra-layer
coupling is weak. The resulting Hamiltonian of the system is \cite{Laubscher-19}
\begin{eqnarray}
\label{Ham_NW}
&&\mathcal{H}_{NW}=\left(k_x^2/2m-2t_z\cos k_z-\mu\right)\tau_z-\alpha k_x\sigma_zs_z\\
&&+\beta\sin k_z\sigma_zs_x+\Gamma\sigma_x\tau_z+\left(\Delta+\Delta_c\cos k_z\right)\sigma_z\tau_ys_y.\nonumber
\end{eqnarray}
Here, $\alpha$ is the Rashba SOC, $\Delta$ is the s-wave
superconducting coupling, $\Gamma$ is the inter-layer tunneling,
$t_z$, $\beta$ is the intra-layer spin-conserving and spin-flipping
hopping terms, correspondingly, and $\Delta_c$ is the cross-Andreev
superconducting coupling.

In the cylinder geometry the examined system is the helical TSC
with the gapless Majorana modes propagating along the edges. Applying
the in-plane magnetic field to this system results in breaking the
time-reversal symmetry and opening the gap in the edge excitation
spectrum in a manner similar to the previously described one for the
helical TSC with the magnetic field. This immediately leads to the
presence of two MCMs in the opposite corners of the two-layered Rashba
nanowire system.

\paragraph{Dimerized monolayer Rashba nanowire system.}
Another way to combine Rashba nanowires was proposed in
\cite{Franca-19}. The wires form a monolayer with
equal spaces between the wires, leading to equal hopping (and
spin-orbital) terms. The dimerization of the system is introduced
with the superconducting phase (supposed to be proximity-induced), with
every second wire having a phase shift. This system was found to be
topologically nontrivial for a wide range of superconducting
phase shifts $\phi$, providing four MCMs in the case of the in-plane
magnetic field and two MCMs in the case of the magnetic field
perpendicular to the plane. In the latter case, while the "magnetic
field" notation is used, one has to introduce it through
ferromagnetic atoms deposited on the superconductor instead of
the external magnetic field since the latter can simultaneously destroy
superconductivity and complicate phase relationships.

\subsection{Braiding on 2D HOTSC}
\label{sec:36}

One of the possibilities to provide quantum computations is
the braiding of anions \cite{nayak-08}. Anions are the particles or
quasi-particles, whose two-particle wave function after their
exchange has the phase different from $0$ or $\pi$ (corresponding
to bosons and fermions, correspondingly). At the moment the Majorana excitations (localized Majorana zero modes expressed in Majorana operators) are the most promising examples of anions. The exchange of two Majorana
fermions leads to the phase shift of $\pm\pi/2$. Furthermore,
the braiding process can be carried out only in the 2D system \cite{nayak-08}; thus,
2D HOTSC seems to be a perspective system for this process.
\begin{figure*}
  \includegraphics[width=0.95\textwidth]{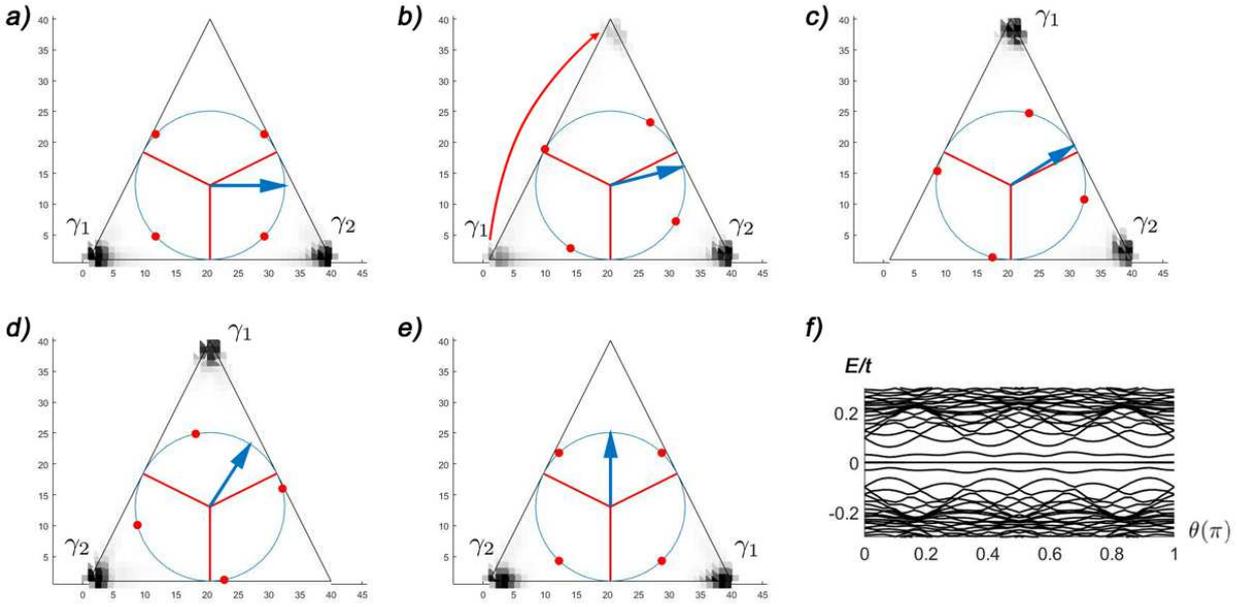}
\caption{Braiding of Majorana corner modes (MCMs) in the triangle-shaped quantum spin-Hall insulator with the $s$-wave superconducting coupling and external magnetic field realized by the rotation of the magnetic field direction (blue arrow). (a) At the starting point, two Majorana modes(MMs) (\ref{phi_QSHI}) with the open disk geometry (red dots) emerges between the normals of two adjacent edges (red lines), providing MCMs at the corresponding corners. Another two MMs fall in one sector between the normals of the adjacent edges, thus causing the corresponding domain wall to vanish and destroying the corresponding MCM. (b-e) The rotation of the in-plane magnetic field forces the MMs positions to cross the normals one after another, thus moving the sector containing two MMs. Hence, the corner without MCMs changes its position, depending on the magnetic field direction and MCMs exchange the positions. Importantly, MCMs hops from one corner to another (b) without spreading along the edge, and thus, the gap in the spectrum remains opened during the process (f).}
\label{fig:PRR2_BR}
\end{figure*}

The braiding process requires changing the parameters of the
system adiabatically in such a way that two Majorana excitations
exchange their position, not intersecting each other (or another
Majorana excitation) and other Majorana excitations must not be
disturbed (in addition, avoiding the creation of new zero-energy
excitations) \cite{Alicea-11}. At the same time, the energy gap between the Majorana
excitations and other excitations must remain sufficient.

These requirements impose constraints on the HOTSC models suitable
for the braiding process. Firstly, the HOTSC model is to provide
the possibility to smoothly control the positions of MCMs in the system.
Obviously, this cannot be done in the systems, whose edge
excitation spectrum structure is determined only by the intrinsic
atomic structure of the system. Secondly, the Majorana excitations
are not allowed to be located at the same position. Consequently, the
systems, in which two Majorana excitations appear in one corner,
cannot be used for braiding. Thirdly, the zero-energy excitations
must remain gapped from other excitations during the process. The
latter prevents the braiding procedure in $p\pm ip$ TSC with
the in-plane magnetic field \cite{Zhu-18} in spite of the fact that the Majorana
excitations can be exchanged by the rotation of the magnetic field
and they are protected from collision by the inverse symmetry.

\begin{figure}
  \includegraphics[width=0.45\textwidth]{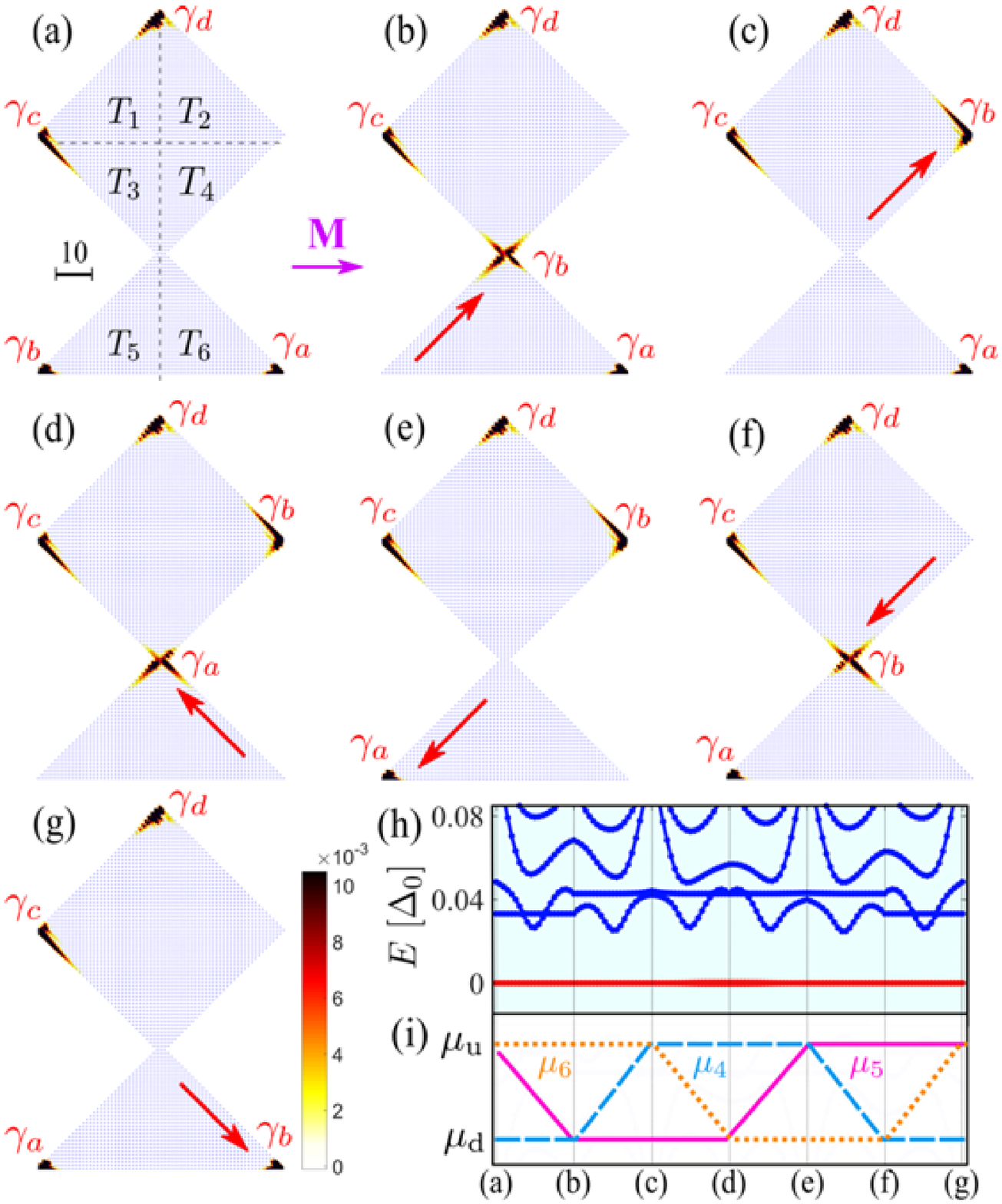}
\caption{Braiding of Majorana corner modes in the device consisting of six triangle shaped islands of higher-order topological superconductors \cite{Zhang-20-prb}. (a-g) Movement of Majorana corner modes during the procedure. (h) Energy spectrum of the system during the whole procedure. (i) Variation of $\mu_{4-6}$ providing braiding.}
\label{fig:PRB_102_BR}
\end{figure}

A rather good example of the braiding process was given in \cite{Pahomi-20}. The proposed HOTSC was constructed from $p\pm ip$ TSC by introducing an additional singlet $s+d$
superconducting coupling and external magnetic field. The proposed
complicated braiding cycle containing a smooth tuning of the
magnetic field direction, value and $s+d$ superconducting coupling
terms prevents the edge spectrum gap from closing during the whole
braiding process and protects MCMs.

A simpler braiding process was described in \cite{Zhang-20-prr}.
While the proposed HOTSC model provides four Majorana excitations
in the open disk (and square) geometry, one can reduce their
quantity to one pair by using the triangle geometry. In this case
there appear two edges with the same sign of the effective mass and one
edge with the opposite effective mass sign independently of the
magnetic field direction. This provides only two domain walls and
consequently, two MCMs. With the rotation of the in-plane magnetic
field one can move these MCMs from one corner to another (Figure \ref{fig:PRR2_BR}). Along with that only one of the MCMs hops directly from one corner to another at any
moment, leaving the edge spectrum gapped during the whole process.

As was noted earlier, the feature
of the g-factor required in the latter HOTSC is necessary only to control the position of the
MMs by the direction of the magnetic field. But as the angle between MMs in one pair in the open disk geometry depends on the ratio between the magnetic field value and the
chemical potential, this position can also be controlled even without the rotation of the magnetic field, though in a restricted manner. By changing the mentioned values, it is possible to move one MCM from one corner to another in the triangle-shaped geometry. Obviously, the braiding
process cannot be performed, as it is impossible to move MMs without the intersection along the whole boundary in this way. However, it is possible to construct a combination of
triangle islands with independently controlled chemical potentials
(gate voltages) in such a way (Figure \ref{fig:PRB_102_BR}) that two Majorana excitations can be
exchanged by moving them from one island to another \cite{Zhang-20-prb}. Remarkably, the proposed braiding procedure can
be done using the electric fields only.

\subsection{Experimental realization and problems of HOTSC}
\label{sec:38}

There are several suggestions for experimental realization.
Concerning TI with the SC coupling as the base of 2D HOTSC,
possible experimental realizations are HgTe/CdTe, InAs/GaSb
quantum wells or magnetic TI such as WTe$_2$ or PbS with the
proximity-induced superconductivity. In the case of the extended s-wave
superconductivity, its origin can be a Fe-based superconductor
(e.g. FeTe$_{1-x}$Se$_x$ monolayer). For the d-wave
superconducting coupling, Cu-based high temperature
superconductors can be used. While the proposed proximity-induced
HOTSCty can be constructed in general, direct calculations of
the proximity effect shows the deviation of the results from the
those predicted in simple models \cite{Li-21}. Instead of using the
superconducting proximity effect, one can also rely on the
intrinsic superconductivity provided by the electron-electron
interactions in TI \cite{Hsu-20,Kheirkhah-20}.

Another possible realization implies the construction of more
complicated heterostructures. For example, for HOTSC
constructed from TI with the Rashba spin-orbit coupling, the device
will consist of two Rashba layers independently connected to the bulk s-wave
superconductors with the superconducting phase shift \cite{Volpez-19} and separated by a dielectric layer. The dielectric layer restricts the tunneling between
Rashba layers and act as an origin of Rashba field in such a
system. In the case of the nanowire-based HOTSC, the proposed device consists of
the nanowires grown on the pairwise connected stripes of the s-wave
superconductor with the phase shift between the pairs or the pairs of
nanowires grown between two s-wave superconductors with the $\pi$
phase shift.

It should be noted that the realization of 2D HOTSC is seriously restricted. The proposed HOTSC models have a square lattice (one can also
find the suggestion for the hexagonal lattice). But there are
no suggestions for HOTSC based on triangle-lattice systems, while
they are known to provide the first-order TSCty. The $C_3$-symmetric systems are considered to be
forbidden to provide the HOTSC phases \cite{Roberts-20,Miert-20}.

\begin{figure}
  \includegraphics[width=0.45\textwidth]{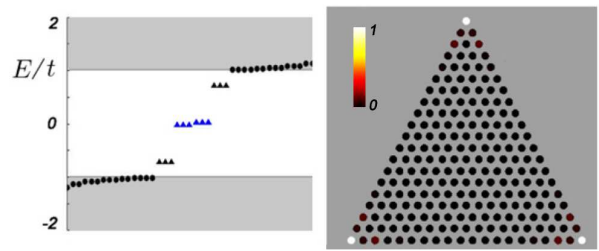}
\caption{Energy spectrum and corner excitation distribution in the triangle-shaped topological insulator with the chiral $d+id$ superconductivity on the triangular lattice \cite{Fedoseev-20}.}
\label{fig:JPCM_32_tr}
\end{figure}

It is still possible to construct a 2D $C_3$
-symmetric superconducting system with corner-localized in-gap
excitations (even zero-energy ones). Following the same procedure performed for TI with the square lattice, it is possible to take a two-band system on the triangular lattice with the inter-band Rashba SOC, which provides the topological insulator phase in this system. Then, one can introduce the $d+id$ superconducting coupling corresponding to the triangular lattice and obtain corner excitations in this system \cite{Fedoseev-20}. In the triangle-shaped geometry there are three excitations (one for each corner) in the system corresponding to the $C_3$-symmetry (Figure \ref{fig:JPCM_32_tr}). They are in-gap for a wide range of the parameters, well localized, and found to be robust against small defects and rather sufficient disorder. One can even control these corner excitations with the magnetic field. However, the energy of these excitations is not pinned to zero, and the excitations are not topologically protected; thus, the examined system is not HOTSC.

Another problem arises from the fact that the appearance of
MCMs in HOTSC depends on the directions of the adjacent edges and
consequently, on the geometry of the sample. As a consequence, the edge
defects can influence the realization of MCMs. Small defects
(or roughness of the edge) neither destroy MCMs nor create new
pairs of MMs in the system, but they only affect the gap size
\cite{Ikegaya-21}. However, if the size of the defect is large
enough and the geometry of the defect generates new domain walls,
new undesired MMs will appear. Such defects make the system
useless for braiding since the uncontrolled creation and annihilation of
the Majorana excitations break the braiding process. The
same problem can be caused by the defects inside the system, as
they generate an additional edge with its own domain
walls and MCMs.

Finally, the main problem of HOTSCty to date is that neither HOTSC
nor HOTI is known among electronic systems. The experimental
realizations of the second-order topological systems in 2D were
mostly obtained in photonic \cite{Hassan-19,Chen-19}, acoustic \cite{Ni-19,Xue-19} and topoelectric systems \cite{Imhof-18,Serra-Garcia-19,Bao-19}. Nevertheless,
the advantages of 2D HOTSCs are promising enough to continue
searching for new ideas for their experimental
realization.

\section{Summary}\label{sec:5}

In the introduction, the general ideas concerning the Majorana modes in one-dimensional and two-dimensional systems are discussed including the well-known spin-orbit coupled quantum wires.

The second section of the review is devoted to different materials and structures with the coexisting spin-singlet superconductivity and noncollinear magnetic ordering which induce the nontrivial topological order in the absence of the spin-orbit interaction. The mentioned magnetic structures include commensurate helical (spiral) textures, particularly 120$^{\circ}$ spin ordering, and cycloidal textures. We discuss the ideas concerning the effective triplet pairings and odd fermion parity of the ground state which provide a rather simple description of the topologically nontrivial phases. The connection between the noncollinear magnetic order and spin-orbit interaction in a uniform magnetic field is given for the 1D and 2D cases. The description of different topological invariants for simultaneously magnetic and superconducting systems is provided. The considered features of the magnetic superconductors support the formation of the Majorana modes at the edges in the strip (or cylinder) geometry and at different defects in the 2D case.

In the third section of the review, the main ideas of the realization of the Majorana bound states on magnetic skyrmions (MSs) are considered. Skyrmions are topologically nontrivial field configurations which are the solutions of nonlinear differential equations. Initially, these solutions were considered by T. Skyrma in nuclear physics for the baryon field. Subsequently, similar field distributions $\textbf{m(r)}$ were found in magnetic systems and liquid crystals. At present, MSs are experimentally found and considered as promising systems for prospective logic and memory devices. The main practical interest in MSs is their locality and topological stability. The latter means the impossibility to convert the skyrmion state into the topologically trivial one without overcoming a very high energy barrier. These properties have recently attracted fundamental interest to the problem of hosting the Majorana modes on MSs. Actually, the developed technologies for controllable movement of MSs make braiding of the Majorana modes and creating stable qubits essentially possible by using MSs. Currently, 2D superconductor / chiral magnet bilayers are most often considered as physical systems in which MSs hosting Majorana modes can be realized. Both superconductors in the Meissner state and with vortices, as well as MSs with complex morphology are considered.

In the fourth section, a brief revision of higher-order topological superconductors (HOTSCs) is provided. Being a novel class of the topologically nontrivial system, HOTSC contains both bulk and edge gapped spectra and provides the Majorana modes on the higher-order boundaries: corners in 2D HOTSC as well as hinges and vertices in 3D HOTSC. The typical way to construct 2D HOTSC is to take a system with a gapless edge states and add a perturbation, which opens the gap in the edge spectrum with the Dirac mass of the opposite sign for the adjacent edges. The corner in this case becomes a domain wall, providing zero energy modes in the way similar to the boundary between two conventional topological insulators with different topological index. While the general idea is the same, the specific method of constructing HOTSC can be rather different. It can be a topological insulator or semimetal with the superconducting coupling of intrinsic or extrinsic character or a TSC perturbed with the magnetic field or magnetic ordering. The properties of HOTSC can be not only bulk-determined but also boundary-determined in contrast to the conventional TSC. In recent investigations, 2D HOTSCs are shown to be perspective systems for the realization of the braiding procedure. The absence of experimental realizations of HOTSC still remains the main problem of HOTSC. But the advantages of 2D HOTSC are perspective enough to continue their investigation.

In general, this review covers a rather wide field of study of topological superconductivity in two-dimensional systems.

\begin{acknowledgements}
We acknowledge fruitful discussions with V.V. Val'kov and S.V. Aksenov. M.S.S. also thanks M.N. Potkina for the valuable discussions. The study was funded by the Russian Foundation for Basic Research (Project No. 19-02-00348), Government of Krasnoyarsk Territory, Krasnoyarsk Regional Fund of Science (Grants No. 19-42-240011, 20-42-243001). A.O.Z. and M.S.S. are grateful to the support of the Foundation for the Advancement of Theoretical Physics and Mathematics ``BASIS''.
\end{acknowledgements}

%
\section*{Conflict of interest}

The authors declare that they have no conflict of interest.

\bibliographystyle{spphys}       
\bibliography{review_JSNM_2021_b}   

%
%

\end{document}